\documentclass[]{interact}
\usepackage{environ} 
\usepackage[dvipsnames]{xcolor}
\usepackage{hyperref}
\usepackage{graphicx}
\usepackage[normalem]{ulem}
\graphicspath{{./gfx/}}

\usepackage{verbatim}

\usepackage{epstopdf}

\usepackage[font=small,labelfont=bf]{caption}
\usepackage{subcaption}
\DeclareCaptionFormat{sqa}{#1~#3}
\usepackage[authoryear]{natbib}

\usepackage{tikz}
\usetikzlibrary{arrows.meta}
\usetikzlibrary{calc}
\usetikzlibrary{patterns}

\usepackage{mathrsfs}
\usepackage{bbm}
\usepackage{bm}

\newcommand{\Pb}{{\mathsf{P}}} 
\newcommand{\Qb}{{\mathsf{Q}}} 
\newcommand{\Eb}{{\mathsf{E}}} 

\DeclareMathOperator{\ARL}{\mathsf{ARL}}   
\DeclareMathOperator{\ARLFA}{\mathsf{ARL2FA}}   
\newcommand{\ESADD}{{\mathsf{ESEDD}}}
\newcommand{\SADD}{{\mathsf{SEDD}}}
\newcommand{\EDD}{{\mathsf{EDD}}}

\newcommand{\mrm}[1]{\mathrm{#1}}
\newcommand{\drm}{{\mrm{d}}}

\newcommand{\mc}[1]{\mathcal{#1}} 

\newcommand{\Ac}{{\mc{A}}}

\newcommand{\mbb}[1]{\mathbb{#1}} 

\newcommand{\class}{{\mbb{C}}}

\newcommand{\mb}[1]{\mathbf{#1}} 

\newcommand{\Xb}{{\mb{X}}}

\newcommand{\bmu}{\mbox{\boldmath $\mu$}}
\newcommand{\btheta}{\mbox{\boldmath $\theta$}}

\newcommand{\set}[1]{\left\{#1\right\}}

\newcommand{\brcs}[1]{\left[#1\right]}

\newcommand{\esssup}{\operatornamewithlimits{ess\,sup}}

\newcommand{\argmax}{\operatornamewithlimits{arg\,max}}

\newcommand{\ignore}[1]{} 

\newcommand{\Fc}{{ \mathscr{F}}} 

\newcommand{\alpham}{{\alpha_{\rm max}}}

\NewEnviron{notforprint*}{\BODY}
\NewEnviron{trivial}{\BODY}
\NewEnviron{ignore*}{}
\NewEnviron{skipproof}{}
\NewEnviron{noignore*}{\BODY}

\newcommand{\nolabel}[1]{}

\hypersetup{
    colorlinks=true,
    bookmarksnumbered=true,
    bookmarksopen=true,
    linkcolor=blue,
    citecolor=blue,
    urlcolor=blue,
    unicode=true,
    breaklinks=true
}
\renewcommand{\geq}{\geqslant}
\renewcommand{\leq}{\leqslant}
\renewcommand{\ge}{\geqslant}
\renewcommand{\le}{\leqslant}
\renewcommand{\cap}{\,{\textstyle\bigcap}\,}
\renewcommand{\cup}{\,{\textstyle\bigcup}\,}

\tikzset{>={Latex[length=8pt, width=4pt]}}

\DeclareMathOperator{\Var}{\mathsf{Var}} 

\DeclareMathOperator{\Hyp}{\mathcal{H}}

\newcommand{\cB}{\mathcal{B}}

\theoremstyle{plain}

\newtheorem*{lemma*}{Lemma}
\newtheorem{theorem}{Theorem}
\newtheorem*{theorem*}{Theorem}
\newtheorem{corollary}{Corollary}
\newtheorem*{corollary*}{Corollary}

\newtheorem*{proposition*}{Proposition}
\theoremstyle{remark}
\newtheorem{remark}{Remark}

\newtheorem*{assumption*}{Assumption}
\theoremstyle{definition}

\newtheorem*{definition*}{Definition}

\DeclareFontFamily{U}{matha}{\hyphenchar\font45}
\DeclareFontShape{U}{matha}{m}{n}{
      <5> <6> <7> <8> <9> <10> gen * matha
      <10.95> matha10 <12> <14.4> <17.28> <20.74> <24.88> matha12
      }{}
\DeclareSymbolFont{matha}{U}{matha}{m}{n}
\DeclareMathSymbol{\abscont}{3}{matha}{"21}

\NewEnviron{revone*}{{\color{Magenta}\BODY~}}
\NewEnviron{revtwo*}{{\color{OliveGreen}\BODY~}}
\NewEnviron{revmisc*}{{\color{ProcessBlue}\BODY~}}

\begin{document}

\title{Change, dependence, and discovery: Celebrating the work of T.~L.~Lai}

\author{
    \name{Alexander~G. Tartakovsky\textsuperscript{a}, Jay Bartroff\textsuperscript{b}, Cheng-Der Fuh\textsuperscript{c}, and Haipeng Xing\textsuperscript{d} 
    \thanks{CONTACT: Jay Bartroff. Email: bartroff@austin.utexas.edu} 
}
    \affil{%
    \textsuperscript{a}AGT StatConsult, Los Angeles, California, USA;
        \textsuperscript{b}University of Texas at Austin, Austin, Texas, USA; 
        \textsuperscript{c}National Central University, Chung-Li, Taiwan;
        \textsuperscript{d}Stony Brook University, NY, USA}
}

\maketitle

\begin{abstract}
Professor Tze Leung Lai made seminal contributions to sequential analysis, particularly in sequential hypothesis testing, changepoint detection and nonlinear renewal theory. His work established fundamental optimality results for the sequential probability ratio test and its extensions, and provided a general framework for testing composite hypotheses. In changepoint detection, he introduced new optimality criteria and computationally efficient procedures that remain influential. He applied these and related tools to problems in biostatistics. In this article, we review these key results in the broader context of sequential analysis. 
\end{abstract}


\newpage
\tableofcontents
\newpage

\section{INTRODUCTION} \label{sec:intro}

This article provides an overview of Tze Leung Lai's major contributions to theoretical statistics, biostatistics, applied probability, 
and sequential analysis. His pioneering work in sequential analysis spans hypothesis testing, changepoint 
detection, and nonlinear renewal theory. He established fundamental optimality results for the sequential 
probability ratio test and its extensions, and developed a general framework for testing composite hypotheses. 
In changepoint detection, he introduced new optimality criteria and computationally efficient procedures that 
continue to shape the field. He applied many of these and related  tools to problems in biostatistics.  We review these advances in the broader context of sequential analysis.

To begin, and to offer a sense of Lai’s remarkable life 
and personality, we present a brief biography.

\subsection{Biographical Sketch}\label{sec:bio}

Tze Leung Lai (June 28, 1945 — May 21, 2023) was a pioneering statistician whose influential research profoundly advanced sequential analysis, stochastic modeling, and statistical decision theory. Born in Hong Kong, he earned his B.A. in Mathematics from the University of Hong Kong in 1967 and his Ph.D. in Statistics from Columbia University in 1971 under the supervision of David Siegmund. His doctoral work on sequential estimation and asymptotic optimality laid the foundation for a distinguished career that combined deep theoretical insight with broad practical impact. In recognition of his early and sustained contributions to the field, he received the COPSS Presidents’ Award in 1983 -- one of the highest honors in statistics.

Lai made seminal contributions to sequential analysis, extending Wald’s classical framework to adaptive, dependent, and high-dimensional settings. He developed asymptotically optimal stopping rules, sequential confidence procedures, and generalized likelihood ratio (GLR) tests that remain fundamental in modern sequential inference. His unified treatment of sequential testing, estimation, and changepoint detection established enduring theoretical principles that continue to influence online learning, sequential experimentation, and real-time statistical monitoring.

At Stanford University, where he served as the Ray Lyman Wilbur Professor of Statistics, Prof.~Lai played a central role in advancing interdisciplinary research. He co-founded and co-directed the Financial and Risk Modeling Institute, where he promoted methodological innovation in financial econometrics, risk management, and financial technology (FinTech). He also directed Stanford’s Center for Innovative Study Design, contributing to the development of adaptive and sequential methods in biostatistics, clinical trials, and health data analysis. His work exemplified the synthesis of rigorous statistical theory with practical applications across science, engineering, and finance.

\begin{figure}[h]
\centering
\includegraphics[height=4.5cm]{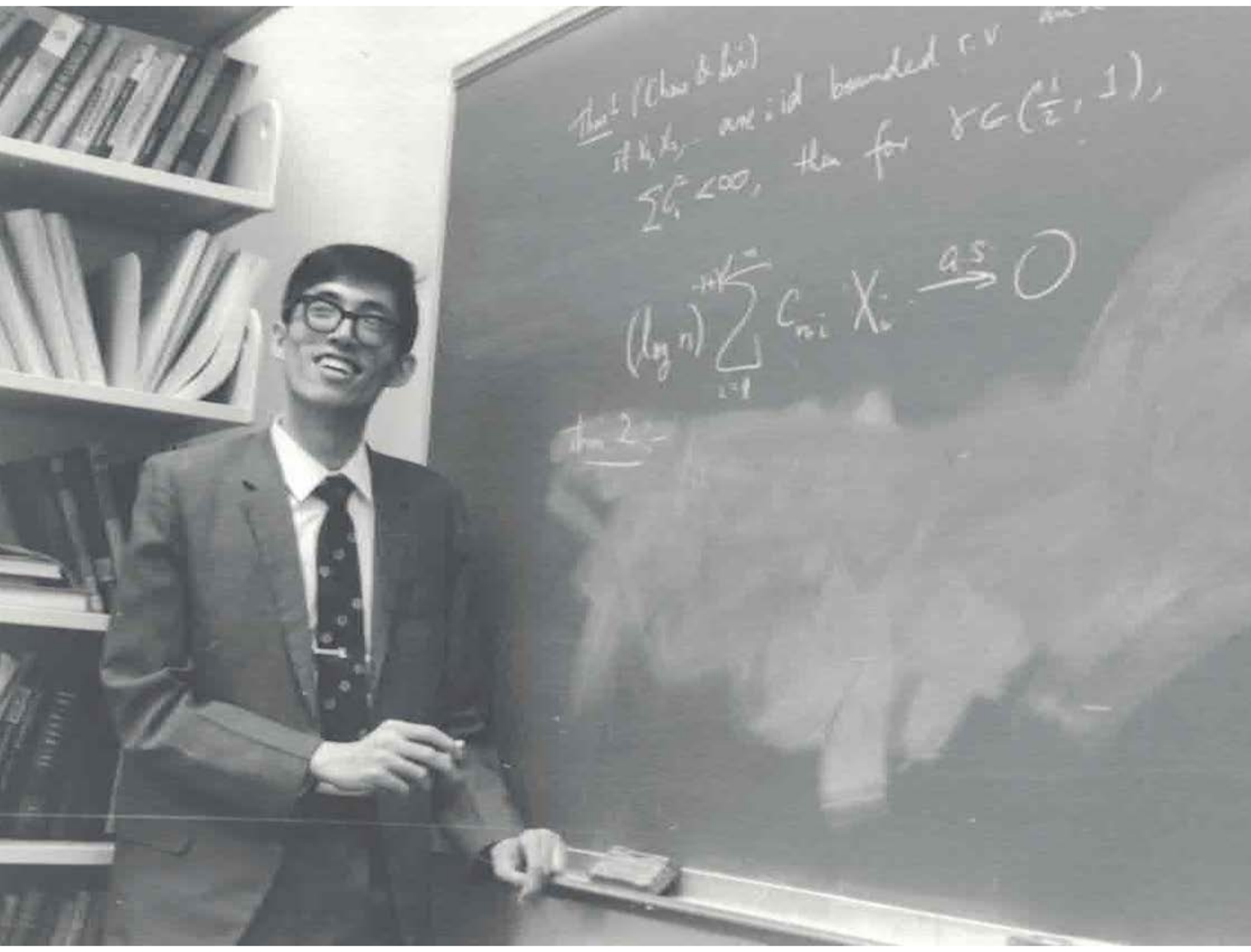} \hspace{0.2cm} 
\includegraphics[height=4.5cm]{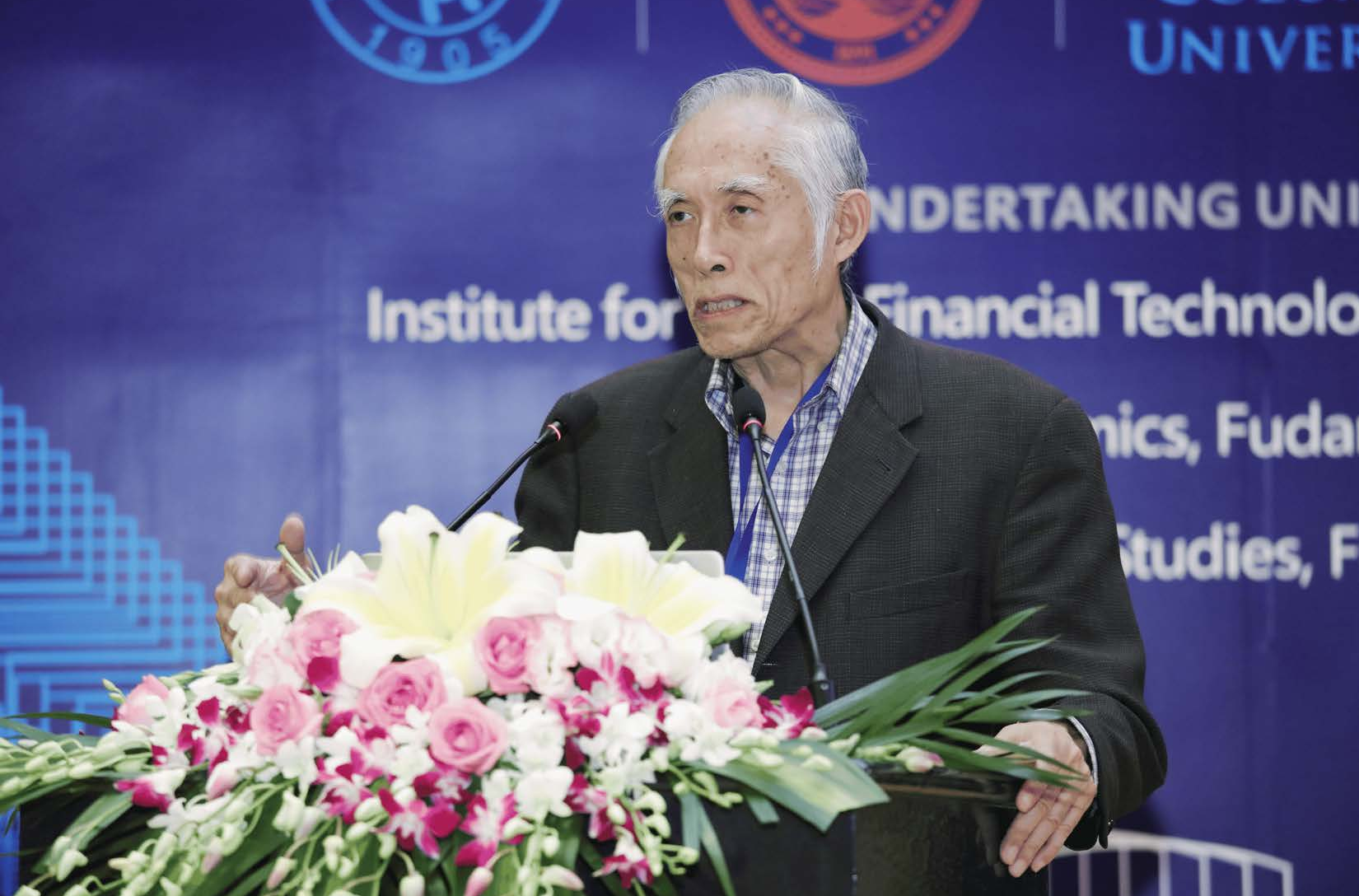}
\caption{Left: Lai in his office at Columbia University, early 1970s. 
Right: Lai presenting a talk at the IMS-FIPS meeting, Shanghai, June 2019.}\label{fig:lai.pics}
\end{figure}

Throughout his remarkable career, Prof.~Lai published more than 300 papers and mentored numerous students and collaborators worldwide. His innovative research reshaped sequential analysis, extending it from its classical foundations into a versatile framework for adaptive learning and data-driven decision-making under uncertainty. He supervised 79 doctoral students and 7 postdoctoral researchers, leaving a lasting legacy of scholarship, mentorship, and inspiration. Figure~\ref{fig:lai.pics} shows Prof.~Lai at two stages of his career.

Prof.~Lai married Letitia Chow in 1975. He is survived by Letitia, their two sons, Peter and David, and two grandchildren. He will be remembered for his profound intellect, generosity of spirit, and enduring influence on the field of modern statistics.

\subsection{The Remainder of this Article}

The remainder of this article highlights Lai’s influential contributions to sequential hypothesis testing, changepoint detection, and nonlinear renewal theory, as well as selected extensions that continue to inspire new developments.

Section~\ref{sec:SHT} reviews sequential hypothesis testing, covering the near-optimality of the SPRT for two simple hypotheses and Lorden’s 2-SPRT in the modified Kiefer--Weiss problem for general non-i.i.d.\ settings, as well as the Bayesian and uniform asymptotic optimality of the GLR test with time-varying boundaries for composite hypotheses.

Section~\ref{sec:CPD} surveys Lai’s contributions to changepoint detection and detection--isolation for non-i.i.d.\ models.

Section~\ref{sec:MCPS} discusses Lai’s work on the sequential analysis of multiple change points. Extending Shiryaev’s Bayesian framework, he developed detection rules for unknown pre- and post-change parameters and for multiple change occurrences. Using hidden Markov models with conjugate priors, his methods recursively compute posterior probabilities of change points, enabling efficient detection and estimation.

Section~\ref{Sec:NRT} reviews Lai’s foundational results on sequential analysis and nonlinear renewal theory, which provide a unified framework for analyzing boundary-crossing times in stochastic processes---central objects in sequential analysis and related statistical applications. We emphasize asymptotic approximations for stopping times and boundary-crossing probabilities.

Finally, Section~\ref{sec:biostat} reviews some of Lai’s work on biomedical clinical trials, including optimal group sequential designs (Section~\ref{sec:GS}) and adaptive designs for mid-trial sample-size re-estimation (Section~\ref{sec:SS.adj}).

\section{SEQUENTIAL HYPOTHESIS TESTING}\label{sec:SHT}

In this section, we review Lai’s contributions to hypothesis testing, particularly his work on establishing the asymptotic optimality of Wald’s SPRT for general non-i.i.d.\ models, and on the near-optimality of the generalized likelihood ratio test for i.i.d.\ exponential families in both Bayesian and frequentist settings \citep{Lai-AS73, Lai-as81-SPRT, Lai-AS1988, LaiZhang-SQA1994}.

\subsection{Near Optimality of Wald's SPRT for Non-i.i.d.\ Models}\label{ssec:SPRT}

We begin with the two-hypothesis testing problem for general non-i.i.d.\ observation models, as addressed by \citet{Lai-as81-SPRT}. Let $(\Omega, \Fc, \{\Fc_n\}, \Pb)$, 
$n \ge 0$, be a filtered probability space, where the sub-$\sigma$-algebra $\Fc_n = \sigma(\Xb^n)$ of~$\Fc$ is generated by the sequence of random variables 
$\Xb^n = \{X_t : 1 \le t \le n\}$ observed up to time~$n$, and $\Fc_0$ is taken to be trivial. The goal is to test a simple null hypothesis $\Hyp_0\!:~ \Pb = \Pb_0$ 
versus a simple alternative $\Hyp_1\!:~ \Pb = \Pb_1$, where $\Pb_0$ and $\Pb_1$ are given probability measures assumed to be locally mutually absolutely continuous. That is, 
their restrictions $\Pb_0^{n} = \Pb_0|_{\Fc_n}$ and $\Pb_1^{n} = \Pb_1|_{\Fc_n}$ to $\Fc_n$ are equivalent for all $1 \le n < \infty$. Let $\Qb^{n}$ denote the restriction of 
a non-degenerate $\sigma$-finite measure $Q$ on $(\Omega, \Fc)$ to $\Fc_n$.

A sequential test is a pair~$\delta = (T, d)$, 
where $T$ is a stopping time with respect to the filtration $\{\Fc_n\}_{n \ge 1}$, and $d = d(\Xb^T)$ is an $\Fc_T$-measurable terminal decision function 
taking values in the set $\{0,1\}$. Specifically, $d = i$ means that hypothesis~$\Hyp_i$ is accepted upon stopping, i.e., 
$\{d = i\} = \{T < \infty,~ \delta~ \text{accepts}~\Hyp_i\}$. Let $\alpha_0(\delta) = \Pb_0(d = 1)$ denote the Type~I error probability (false positive), and 
$\alpha_1(\delta) = \Pb_1(d = 0)$ denote the Type~II error probability (false negative) of the test~$\delta$.

For $i=0,1$, define the class of tests $\class(\alpha_0, \alpha_1) = \left\{ \delta : \alpha_i(\delta) \le \alpha_i, ~ i=0,1 \right\}$,
that is, the set of procedures whose error probabilities $\alpha_i(\delta)$ do not exceed the prescribed thresholds $0 < \alpha_i < 1$.

Consider the general non-i.i.d.\ model, where the observed random variables $X_1, X_2, \dots$ may be dependent and non-identically distributed. Under $\Pb_i$, 
the sample $\Xb^n = (X_1, \dots, X_n)$ has a joint density~$p_i(\Xb^n)$ with respect to the dominating measure $\Qb^n$ for all $n \ge 1$, which can be expressed as
\begin{equation}\label{Jointdensnoniid}
p_i(\Xb^n) = \prod_{t=1}^n f_{i,t}(X_t \mid \Xb^{t-1}), \quad i = 0, 1,
\end{equation}
where $f_{i,t}(X_t \mid \Xb^{t-1})$ denotes the conditional density of $X_t$ given the set of past observations $\Xb^{t-1}$ under hypothesis~$\Hyp_i$.

For $n\ge 1$, define  the log-likelihood ratio (LLR) process between the hypotheses $\Hyp_1$ and~$\Hyp_0$ as
\begin{align*}
\lambda_n & = \log \frac{\drm \Pb_1^{n}}{\drm \Pb_0^{n}}(\Xb^n) =  \log \frac{p_{1}(\Xb^n)}{p_{0}(\Xb^n)}  = \sum_{t=1}^n \log \brcs{\frac{f_{1,t}(X_t \mid \Xb^{t-1})}{f_{0,t}(X_t\mid \Xb^{t-1})}} .
\end{align*}

\citet{wald45,wald47} introduced the {\em Sequential Probability Ratio Test} (SPRT). 
Letting $a_0 < 0$ and $a_1 > 0$ be thresholds, Wald's SPRT $\delta_*(a_0, a_1) = (T_*, d_*)$ is defined by
\begin{equation}\label{SPRT}
T_*(a_0, a_1) = \inf\left\{n \ge 1 : \lambda_n \notin (a_0, a_1)\right\}, \quad
d_*(a_0, a_1) = 
\begin{cases}
1 & \text{if}~ \lambda_{T_*} \ge a_1, \\
0 & \text{if}~ \lambda_{T_*} \le a_0.
\end{cases}
\end{equation}

For $i = 0, 1$, let $\Eb_i$ denote expectation under hypothesis~$\Hyp_i$, i.e., under~$\Pb_i$.

In the i.i.d.\ case, when \( f_{i,t}(X_t \mid \Xb^{t-1}) = f_i(X_t) \) in \eqref{Jointdensnoniid}, Wald's SPRT exhibits a remarkable optimality property: it minimizes both expected sample sizes, \( \Eb_0[T] \) and \( \Eb_1[T] \), over the entire class of sequential and non-sequential tests \( \class(\alpha_0, \alpha_1) \),
as established by \citet{wald48}.

However, in the non-i.i.d.\ case—when the log-likelihood ratio $\lambda_n$ is no longer a random walk—the Wald–Wolfowitz argument breaks down, and their theorem on SPRT optimality 
can no longer be applied. This limitation holds even in certain i.i.d.\ settings, such as invariant sequential tests, which fall outside the scope of their original proof.

This limitation motivated \citet{Lai-as81-SPRT} to extend the Wald–Wolfowitz result by proving that the SPRT is also first-order asymptotically optimal as 
$\alpham = \max(\alpha_0, \alpha_1) \to 0$ for general 
non-i.i.d.\ models with dependent and non-identically distributed observations, provided that the normalized log-likelihood ratio $n^{-1} \lambda_n$ converges 
$r$-quickly to finite limits $I_i$ under $\Pb_i$, for $i = 0, 1$. 

Given the fundamental importance of this result for both theory and applications, we now present a detailed overview of Lai's 1981 paper.

First, Lai observed that in the i.i.d.\ case, the normalized LLR \( n^{-1} \lambda_n \) converges almost surely (a.s.) to \( I_1 > 0 \) under \( \Pb_1 \), and to \( I_0 < 0 \) 
under \( \Pb_0 \), where \( I_i = \Eb_i[\lambda_1] \), \( i = 0,1 \).

This observation naturally led to the idea that, in the general non-i.i.d.\ case, one should assume the following condition:
\begin{equation}\label{ASconvLLR}
\frac{1}{n} \lambda_n \xrightarrow[n \to \infty]{\Pb_i\text{-a.s.}} I_i, \quad i = 0, 1,
\end{equation}
where the limits $I_0 < 0$ and $I_1 > 0$ are finite.

This allowed Lai to prove the following \emph{weak asymptotic optimality} result: if the almost sure convergence condition~\eqref{ASconvLLR} holds and the SPRT $\delta_* \in \class(\alpha_0, \alpha_1)$ has thresholds $a_0 \sim \log \alpha_1$, $a_1 \sim |\log \alpha_0|$ as $\alpham = \max(\alpha_0,\alpha_1) \to 0$, then for $i=0,1$ and every $\varepsilon \in (0,1)$,
\[
\inf_{\delta \in \class(\alpha_0, \alpha_1)} 
\Pb_i\!\left(T > \varepsilon\, T_*(a_0,a_1)\right) \;\to\; 1 
\quad \text{as } \alpham \to 0.
\]
See Theorem~1 in \citet{Lai-as81-SPRT}.

However, this result is not entirely practical. In applications, one is typically interested in the expectation of the stopping time, or more generally, in its higher moments. 
In fact, almost sure convergence alone does not even guarantee the finiteness of the expected sample size under general non-i.i.d.\ models. 

To address this, Lai used the concept of \emph{$r$-quick convergence} of the log-likelihood ratio process to establish near-optimality with respect to moments of the stopping 
time distribution. The definition used in our context is as follows.

For $i=0,1$, $\varepsilon > 0$ and $n \ge 1$, define the last exit times
$\tau_{i,\varepsilon} = \sup \left\{ n \ge 1 : \left| \lambda_n - n I_i \right| > \varepsilon \right\}$, where $\sup\{\varnothing\} =0$.
For $r > 0$, the LLR process is said to converge \emph{$r$-quickly} to $I_i$ under $\Pb_i$ if
\begin{equation}\label{rquickLLR}
\Eb_i[\tau_{i,\varepsilon}^r] < \infty \quad \text{for all } \varepsilon > 0, \quad i = 0, 1.
\end{equation}

The following theorem establishes the first-order asymptotic optimality of the SPRT with respect 
to $r$th moments under the $r$-quick convergence conditions \eqref{rquickLLR} for the normalized 
LLR process $\lambda_n/n$. This is the outstanding result of Lai's 1981 paper.

\begin{theorem}[SPRT Asymptotic Optimality]\label{Th:AOSPRTnoniid}
Let $r > 0$. Assume that there exist finite constants $I_0 < 0$ and $I_1 > 0$ such that $n^{-1} \lambda_n$ converges $r$-quickly to $I_i$ under $\Pb_i$ for $i = 0, 1$, i.e., the conditions \eqref{rquickLLR} are satisfied. Then, as $\alpham \to 0$,
\begin{equation}\label{Asymptnoniid}
\begin{split}
\inf_{\delta \in \class(\alpha_0, \alpha_1)} \Eb_0[T^r] & \sim \left( \frac{|\log \alpha_1|}{|I_0|} \right)^r \sim \Eb_0[T_*^r], \\
\inf_{\delta \in \class(\alpha_0, \alpha_1)} \Eb_1[T^r] & \sim \left( \frac{|\log \alpha_0|}{I_1} \right)^r \sim \Eb_1[T_*^r].
\end{split}
\end{equation}
\end{theorem}

The results of Theorem~\ref{Th:AOSPRTnoniid} can be extended to the asymptotically non-stationary case when the LLR process is normalized by a non-linear function $\psi(n)$
that goes to infinity and increases faster than the logarithmic function, for example, a polynomial function $\psi(t) = t^k$, $k > 0$. Specifically, if we require that 
\[
\frac{\lambda_n}{\psi(n)} \xrightarrow[n \to \infty]{\Pb_i\text{-}r\text{-quickly}} I_i, \quad i = 0, 1,
\]
then in place of the asymptotic expressions \eqref{Asymptnoniid} the following relations hold:
\begin{equation*}
\begin{split}
\inf_{\delta \in \class(\alpha_0, \alpha_1)} \Eb_0[T^r] &\sim \left[ \psi^{-1}\!\biggl(\frac{|\log \alpha_1|}{|I_0|}\biggr) \right]^r \sim \Eb_0[T_*^r], 
\\
\inf_{\delta \in \class(\alpha_0, \alpha_1)} \Eb_1[T^r] &\sim \left[ \psi^{-1}\!\biggl(\frac{|\log \alpha_0|}{I_1}\biggr) \right]^r \sim \Eb_1[T_*^r],
\end{split}
\end{equation*}
where $\psi^{-1}(t)$ denotes the inverse function of $\psi(t)$. The details can be found in \cite{TartakovskySISP98} and \cite[Section~3.4]{TNB_book2014}.

These results also hold under slightly less restrictive conditions for the LLR process. Specifically, when $\lambda_n/\psi(n)$ converges to $I_i$ $r$-completely under probability
measures $\Pb_i$, that is, if
\begin{equation*}
\lim_{n\to\infty} \sum_{t=n}^\infty t^{r-1} \Pb_i\Bigl(\bigl\lvert \tfrac{\lambda_t}{\psi(t)} - I_i \bigr\rvert > \varepsilon\Bigr) = 0 
\quad \text{for every } \varepsilon > 0, \quad i = 0, 1.
\end{equation*}
See \cite{TartakovskyAMSA2025} for details.

\subsection{Near Optimality of Lorden's 2-SPRT for Non-i.i.d.\ Models}\label{ssec:2SPRT}

\citet{lorden-as76} advanced the modified Kiefer--Weiss problem for two hypotheses 
$\Hyp_i:\Pb=\Pb_i$, $i=0,1$, by introducing the 2-SPRT---a combination of one-sided SPRTs---and 
proving its third-order asymptotic optimality as $\alpham \to 0$ under the intermediate measure $\Pb_2$. 
He later described the structure of optimal tests \citep{lorden-ptrf80}. Extending this setting to dependent, 
non-i.i.d.\ observations, \citet{Lai-as81-SPRT} showed that the 2-SPRT is first-order asymptotically optimal 
whenever the normalized LLRs $n^{-1} \lambda_n^{(i)} =  n^{-1} \log[p_2(\Xb^n)/p_i(\Xb^n)]$
converge $r$-quickly to finite limits $\eta_i \ge 0$ under $\Pb_2$, with $\max\{\eta_0,\eta_1\}>0$.

Next, letting $a_0 > 0$ and $a_1 > 0$ be thresholds, Lorden’s 2‐SPRT 
$\delta_2(a_0,a_1) = \bigl(T_2(a_0,a_1),\,d_2(a_0,a_1)\bigr)$ is defined as
\begin{equation}\label{2SPRT}
\begin{split}
T_2(a_0, a_1) &= \inf\Bigl\{\,n \ge 1 : \lambda_n^{(0)} \ge a_0 \;\text{or}\; \lambda_n^{(1)} \ge a_1\Bigr\},  \quad \inf \{\varnothing\}= \infty,
\\
d_2(a_0, a_1) &= 
\begin{cases}
1, & \text{if } \lambda_{T_2}^{(0)} \ge a_0,\\
0, & \text{if } \lambda_{T_2}^{(1)} \ge a_1.
\end{cases}
\end{split}
\end{equation}

It is easily shown that the error probabilities of the 2‐SPRT satisfy the inequalities
\[
\Pb_0\bigl(d_2 = 1\bigr) \;\le\; \exp\set{-a_0} \,\Pb_2\bigl\{d_2 = 1\bigr\}, 
\quad 
\Pb_1\bigl(d_2 = 0\bigr) \;\le\; \exp\set{-a_1} \,\Pb_2\bigl\{d_2 = 0\bigr\},
\]
which implies that setting $a_0 = \log \alpha_0^{-1}$ and $a_1 = \log \alpha_1^{-1}$ ensures the 2‐SPRT belongs to the class $\class(\alpha_0, \alpha_1)$.

The following theorem, which follows from Lai’s Theorem~2 and Corollary~2, 
is the centerpiece of the modified Kiefer--Weiss non‐i.i.d.\ theory.

\begin{theorem}[2-SPRT Asymptotic Optimality]\label{Th:AO2SPRTnoniid}
Let $r > 0$. Assume that there exist finite constants $\eta_0 \ge 0$ and $\eta_1 \ge 0$ with $\max\{\eta_0,\eta_1\} > 0$ such that the normalized log-likelihood ratios 
$n^{-1} \lambda_n^{(i)}$ converge $r$-quickly to $\eta_i$ under $\Pb_2$.
Then, as $\alpham \to 0$,
\begin{equation}\label{Asymptnoniid2SPRT}
\inf_{\delta \in \class(\alpha_0, \alpha_1)} \Eb_2[T^r] 
\;\sim\; 
\biggl[\min\Bigl\{\frac{|\log \alpha_1|}{\eta_1},\,\frac{|\log \alpha_0|}{\eta_0}\Bigr\}\biggr]^r 
\;\sim\; 
\Eb_2\bigl[T_2(\alpha_0,\alpha_1)^r\bigr].
\end{equation}
\end{theorem}

\begin{remark}
Lai's paper was groundbreaking in showing that invariant sequential tests can be asymptotically optimal, 
illustrated through several challenging examples. Its influence, however, extends well beyond invariant tests 
for i.i.d.\ models, as later demonstrated in multi‐decision problems by
\cite{TartakovskySISP98, DTVPart1_IEEEIT1999, TNB_book2014, Tartakovsky_book2020, TartakovskyAMSA2025}.
\end{remark}

\subsection{Near Optimality of GLR Sequential Tests for Composite Hypotheses} \label{ssec:GLR}

For practical purposes, it is preferable to design tests that minimize $\Eb_\theta[T]$ uniformly over all parameters, rather than only at some intermediate, typically least favorable, point. We focus on sequential tests that are approximately uniformly optimal for small error probabilities, or asymptotically Bayesian when the observation cost is small, in the context of composite hypothesis testing

Let $X_1,X_2,\dots$ be i.i.d.\ with density $f_\theta$, $\theta \in \Theta = \Theta_0 \cup \Theta_1 \cup \Theta_{\mathrm{in}}$. 
We test $\Hyp_0\colon \theta \in \Theta_0$ vs.\ $\Hyp_1\colon \theta \in \Theta_1$, with $\Theta_{\mathrm{in}}$ an indifference zone. 
The goal is a sequential test $\delta=(T,d)$ minimizing $\Eb_\theta[T]$ uniformly over $\Theta$ within
\begin{equation}\label{classcomp2hyp}
\class(\alpha_0,\alpha_1) 
= \bigl\{\delta : \sup_{\theta \in \Theta_0} \Pb_\theta(d = 1) \le \alpha_0,\ 
\sup_{\theta \in \Theta_1} \Pb_\theta(d = 0) \le \alpha_1 \bigr\}, \quad 0<\alpha_i<1.
\end{equation}

Since a strictly uniformly optimal test does not exist, we consider asymptotics as the error probabilities vanish, in which case first-order asymptotic optimality in the frequentist setting means
\begin{equation}\label{FOopt}
\lim_{\alpham \to 0} 
\frac{\inf_{\delta\in\class(\alpha_0,\alpha_1)} \Eb_\theta[T]}{\Eb_\theta[T]} = 1, 
\quad \theta \in \Theta.
\end{equation}

In the Bayesian setting with prior $\pi(\theta)$, cost $c$, and loss $L(\theta)$, the integrated risk is
\[
\rho_c^\pi(\delta) = \int_{\Theta_0} L(\theta) \Pb_\theta(d=1)\,\drm \pi(\theta) + \int_{\Theta_1} L(\theta) \Pb_\theta(d=0)\,\drm \pi(\theta) + c \int_\Theta \Eb_\theta[T]\,\drm\pi(\theta),
\]
and tests are asymptotically optimal as $c\to 0$:
\begin{itemize}
  \item First‐order: $\inf_\delta \rho_c^\pi(\delta) = \rho_c^\pi(\delta)(1+o(1))$,
  \item Second‐order: $\inf_\delta \rho_c^\pi(\delta) = \rho_c^\pi(\delta)+O(c)$,
  \item Third‐order: $\inf_\delta \rho_c^\pi(\delta) = \rho_c^\pi(\delta)+o(c)$.
\end{itemize}

Consider a single‐parameter exponential family $\{\Pb_\theta : \theta \in \Theta\}$ with densities
\begin{equation}\label{Expfam}
\frac{f_\theta(x)}{f_{\tilde\theta}(x)} = \exp\bigl\{(\theta-\tilde\theta)x - [b(\theta)-b(\tilde\theta)]\bigr\},
\end{equation}
where $b(\theta)$ is convex and smooth on $\widetilde\Theta \subset \Theta$. We test
$\Hyp_0: \underline{\theta}\le \theta \le \theta_0$ vs.\ 
$\Hyp_1: \theta_1 \le \theta \le \overline{\theta}$, with indifference interval $\Theta_{\mathrm{in}} = (\theta_0,\theta_1)$ of positive width.  

The optimal Bayesian test $\delta_{\rm opt}=(T_{\rm opt},d_{\rm opt})$ is
\[
T_{\rm opt} = \inf\{n\ge 1 : (S_n,n) \in \cB_c\}, \qquad
d_{\rm opt}=j \text{ if } (S_n,n)\in \cB_c^j,~ j=0,1,
\]
where $S_n = \sum_{i=1}^n X_i$ and $\cB_c = \cB_c^0 \cup \cB_c^1$ is determined numerically.

\citet{Schwarz-AMS1962} derived the test $\hat{\delta}(\hat\theta)$, where $\hat\theta = \{\hat\theta_n\}$ is the maximum likelihood estimator (MLE) of $\theta$, as an asymptotic solution, as $c \to 0$, to the Bayesian problem with the 0–1 loss function: $L(\theta)=0$ if the decision is correct and $L(\theta)=1$ otherwise.
In this setting, the Bayes integrated risk of a sequential test $\delta$ is
\begin{equation}\label{AvRisk}
\rho_c^\pi(\delta) 
= \int_{\underline{\theta}}^{\theta_0} \Pb_\theta\bigl(d = 1\bigr)\,\drm \pi(\theta)
+ \int_{\theta_1}^{\overline{\theta}} \Pb_\theta\bigl(d = 0\bigr)\,\drm \pi(\theta)
+ c \int_{\underline{\theta}}^{\overline{\theta}} \Eb_\theta\bigl[T\bigr]\,\drm \pi(\theta).
\end{equation}

Denote by
\[
\lambda_n(\theta, \theta_i) 
= \sum_{t=1}^n \log \frac{f_{\theta}(X_t)}{f_{\theta_i}(X_t)} 
= \bigl[\theta\,S_n - n\,b(\theta)\bigr] \;-\; \bigl[\theta_i\,S_n - n\,b(\theta_i)\bigr]
\]
the log-likelihood ratio between the points $\theta$ and $\theta_i$.

Schwarz's test prescribes stopping sampling at 
$
\widehat{T}(\hat\theta) \;=\; \min\bigl\{\widehat{T}_0(\hat\theta),\,\widehat{T}_1(\hat\theta)\bigr\},
$
where, for each $i = 0, 1$,
\begin{equation}\label{hatTiSchwarz}
\begin{aligned}
\widehat{T}_i(\hat\theta) 
& = \inf\biggl\{\,n \ge 1 : \lambda_n\bigl(\hat\theta_n, \theta_i\bigr) \;\ge\; \lvert \log c\rvert\biggr\} , \quad \inf\varnothing = \infty.
\end{aligned}
\end{equation}
The terminal decision rule $\hat{d}(\hat\theta)$ of the test 
$
\hat\delta(\hat\theta) \;=\; \bigl(\widehat{T}(\hat\theta),\,\hat{d}(\hat\theta)\bigr)
$
accepts $\Hyp_0$ if 
$
\hat\theta_{\widehat{T}} < \theta^*,
$
where $\theta^*$ satisfies 
$
I\bigl(\theta^*, \theta_0\bigr) \;=\; I\bigl(\theta^*, \theta_1\bigr),
$
and 
\[
I(\theta, \theta_i) \;=\; \Eb_\theta\bigl[\lambda_1(\theta,\theta_i)\bigr] 
\;=\; (\theta - \theta_i)\,b'(\theta) \;-\; \bigl[b(\theta) - b(\theta_i)\bigr].
\]

Note also that
\begin{equation}\label{SchST}
\widehat{T}(\hat\theta) \;=\; 
\inf\Bigl\{\,n \ge 1 : n\,\max\bigl[I\bigl(\hat\theta_n,\theta_0\bigr),\,I\bigl(\hat\theta_n,\theta_1\bigr)\bigr] 
\;\ge\; \lvert \log c\rvert \Bigr\}.
\end{equation}

A significant advancement in Bayesian theory for testing separated hypotheses about the parameter of the one‐parameter exponential family \eqref{Expfam} was made by \citet{Lorden-unpublished-1977}. \citet{Lorden-unpublished-1977} introduced a family of GLR tests for separated hypotheses in a single‐parameter exponential family. His tests achieve third‐order asymptotic Bayes optimality, $\rho_c^\pi(\widehat\delta) = \inf_\delta \rho_c^\pi(\delta) + o(c)$ as $c \to 0$, by using a slightly reduced threshold and adaptive weight functions correcting for overshoots. These modifications make Lorden’s GLR test nearly optimal compared with the standard Schwartz's GLR.

\citet{Lai-AS1988} improved Bayesian testing of two composite hypotheses by letting the indifference interval $\Delta = \theta_1 - \theta_0$ shrink as the observation cost $c \to 0$, resulting in a GLR test with adaptive, time-varying boundaries. This unified approach covers both tests with an indifference zone, $\Hyp_0: \theta \le \theta_0$ vs.\ $\Hyp_1: \theta \ge \theta_1$, and without, $\Hyp_0: \theta < \theta_0$ vs.\ $\Hyp_1: \theta > \theta_0$.

Due to importance of Lai's theory we now provide details.

In the problem with the indifference zone, Lai proposed replacing the constant threshold $|\log c|$ in Schwarz’s stopping time \eqref{SchST} with a time‐varying boundary $g(cn)$:
\begin{equation}\label{LaiST}
T^*(\hat\theta) 
\;=\; 
\inf\Bigl\{\,n \ge 1 : n\,\max\bigl[I(\hat\theta_n,\theta_0),\,I(\hat\theta_n,\theta_1)\bigr] 
\;\ge\; \lvert \log g(cn)\rvert \Bigr\},
\end{equation}
where $g\colon (0,\infty)\to[0,\infty)$ is any function satisfying conditions detailed below.

If we set $\theta_0 = \theta_1$ in \eqref{LaiST}, the stopping rule simplifies to
\begin{equation}\label{LaiST1}
T^*(\hat\theta) 
\;=\; 
\inf\Bigl\{\,n \ge 1 : n\,I\bigl(\hat\theta_n,\theta_0\bigr)
\;\ge\; \lvert \log g(cn)\rvert \Bigr\}.
\end{equation}

Hence, the stopping rules \eqref{LaiST} and \eqref{LaiST1} provide a unified treatment of both testing problems: separated hypotheses with an indifference zone, and
one-sided hypotheses without an indifference zone.

Recall that the integrated Bayesian risk \(\rho_c^\pi(\delta)\) was defined in \eqref{AvRisk}.

The following theorems summarize Lai’s unified hypothesis‐testing theory. The first addresses asymptotic optimality with an indifference zone, the second without. Let $J(\theta) = \max\{I(\theta, \theta_0), I(\theta, \theta_1)\}$.

\begin{theorem}[Bayesian Optimality with Indifference Zone]\label{Th:LaiIndif}
Consider testing hypotheses with an indifference zone. Let $\xi > -\tfrac12$ and $g:(0,\infty)\to[0,\infty)$ satisfy, as $t\to0$, 
\[
g(t) \sim |\log t|, \qquad g(t) \ge |\log t| + \xi \log|\log t|.
\]
Under suitable prior conditions for fixed $\theta_0 < \theta_1$, the GLR test $\delta^*=(T^*,d^*)$ achieves
\[
\inf_\delta \rho_c^\pi(\delta) \;\sim\; c\,|\log c| \int_{\underline{\theta}}^{\overline{\theta}} \frac{d\pi(\theta)}{J(\theta)} \;\sim\; \rho_c^\pi(\delta^*), \quad c\to0.
\]
If additionally $\pi$ is positive and continuous near $\theta_0$ and $(\theta_1-\theta_0)^2/c \to \infty$, then
\[
\inf_\delta \rho_c^\pi(\delta) \;\sim\; \frac{8\,\pi'(\theta_0)}{b''(\theta_0)}\, \frac{c}{\theta_1 - \theta_0} \log\!\Bigl(\frac{(\theta_1 - \theta_0)^2}{c}\Bigr) \;\sim\; \rho_c^\pi(\delta^*), \quad c\to0.
\]
\end{theorem}

Let $\{w(t)\}_{t\ge0}$ denote a Wiener process with drift $\mu$ under $\Pb_\mu$, and define the stopping time
$
\tau = \inf\{t>0 : |w(t)| \ge h_0(t)\},
$
where $h_0$ is a positive function on $(0,\infty)$ satisfying conditions (2.5) and (2.6) in \citet{Lai-AS1988}.

\begin{theorem}[Bayesian Optimality, No Indifference Zone]\label{Th:LaiNoIndif}
For testing without an indifference zone and a prior $\pi$ positive near $\theta_0$, the GLR test $\delta^*=(T^*,d^*)$ satisfies, as $c\to0$,
\[
\inf_\delta \rho_c^\pi(\delta) \sim \rho_c^\pi(\delta^*) \sim 
\frac{c^{1/2}\pi'(\theta_0)}{(b''(\theta_0))^{1/2}}
\int \bigl\{ \Eb_\mu[\tau] + \Pb_\mu(w(\tau)>0\text{ or }<0) \bigr\}\,d\mu.
\]
\end{theorem}

In addition to these Bayesian results, Lai also established uniform (over all \(\theta\)) asymptotic optimality of his GLR test within the class \(\class(\alpha_0,\alpha_1)\) of tests with prescribed error probabilities, thereby solving the frequentist problem \eqref{FOopt}.  Let
\[
\alpha_0^* \;=\; \sup_{\theta \le \theta_0} \Pb_{\theta}\bigl(d^* = 1\bigr)
\;=\; \Pb_{\theta_0}\bigl(\hat{\theta}_{T^*} > \theta^*\bigr),
\quad
\alpha_1^* \;=\; \sup_{\theta \ge \theta_1} \Pb_{\theta}\bigl(d^* = 0\bigr)
\;=\; \Pb_{\theta_1}\bigl(\hat{\theta}_{T^*} \le \theta^*\bigr).
\]
\begin{theorem}[Uniform Asymptotic Optimality]\label{Th:LaiUniform}
If $|\log \alpha_0^*| \sim |\log \alpha_1^*| \sim |\log c|$, then,
 as $c\to 0$,
\[
\Eb_\theta[T^*] \sim \frac{|\log c|}{J(\theta)} \sim \inf_{\delta\in \class(\alpha_0^*,\alpha_1^*)} \Eb_\theta[T]
\quad \text{uniformly for } \theta \in \Ac.
\]
\end{theorem}

Lai’s adaptive GLR $\delta^*$ uses the MLE $\hat\theta_n$ to adjust the stopping boundary $g(cn)$, achieving near-optimal performance across $\theta$ for both indifference-zone and one-sided hypotheses. Unlike Lorden’s random MLE-based boundaries \citep{Lorden-unpublished-1977}, Lai’s are deterministic and handle both cases.

Later, \citet{LaiZhang-SQA1994} extended these adaptive GLR tests to multiparameter exponential families, addressing one‐sided tests of smooth scalar functions of the vector parameter.

\section{QUICKEST CHANGE DETECTION AND ISOLATION} \label{sec:CPD}

\subsection{Lai's Changepoint Detection Theory for Non--i.i.d.\ Data}\label{ssec:CPDnoniid}

In many applications, data may change distribution at an unknown time. A variety of examples were discussed in \cite{TNB_book2014,Tartakovsky_book2020}.

In sequential changepoint detection, the goal is to detect distributional changes quickly while controlling false alarms—the quickest change (or disorder) detection problem.

The field originated in quality control with Shewhart’s charts \citep{shewhart-book31}, and advanced in the 1950s–70s with the optimal and nearly optimal methods of \citet{GR1952}, \citet{page-bka54}, \citet{lorden-ams71}, \citet{shiryaev-tpa1963, shiryaev-book1969}.

For the classical i.i.d.\ case, let  $X_1, X_2, \dots$be independent with change point \(\nu\), i.e., $X_1,\dots,X_{\nu-1}\sim F_0$, $X_\nu,X_{\nu+1},\dots\sim F_1$, where distributions
 \(F_0,F_1\) have densities \(f_0,f_1\).   A detection rule is a stopping time \(T\) at which a change is declared.

Let  \(\Pb_\nu, \Pb_\infty\) and \(\Eb_\nu, \Eb_\infty\) denote probability measures and expectations when the change point \(\nu\) is fixed (\(1\le \nu < \infty\)) and when $\nu=\infty$ 
(i.e., no change ever occurs). 

\ignore{
Under this model, the joint density can be written as
\begin{equation}\label{iidmodel}
p_{\nu}(\Xb^n) = 
\begin{cases}
\displaystyle
\prod_{t=1}^{\nu-1} f_0(X_t)
\;\times\;
\prod_{t=\nu}^n f_1(X_t), 
& n \ge \nu, 
\\[6pt]
\displaystyle
\prod_{t=1}^{n} f_0(X_t),
& n < \nu.
\end{cases}
\end{equation}
}

Let 
$
Z_t \;=\; \log[f_1(X_t)/f_0(X_t)]
$
denote the LLR for the \(t\)th observation \(X_t\).

The CUSUM test \citep{page-bka54} is defined by
\[
W_n=\max_{1\le\nu\le n}\sum_{t=\nu}^n Z_t,\qquad
T_a=\inf\{n\ge1:W_n\ge a\}, \quad \inf \varnothing = \infty,
\]
where \(a>0\).  Here \(W_n\) is the log–GLR statistic and satisfies
\[
W_n=(W_{n-1}+Z_n)^+,\quad n \ge 1, \quad W_0=0.
\]

\citet{page-bka54} evaluated the CUSUM procedure using the average run length (ARL) to false alarm, 
$\ARLFA(T) = \Eb_\infty[T]$, and the ARL to detection, $\ARL(T) = \Eb_1[T]$, in the case where the change occurs at the very beginning. 
A more informative measure is the conditional detection delay, $\Eb_\nu[T-\nu \mid T>\nu]$,
but under the constraint $\ARLFA(T) \ge \gamma$ there exists no procedure that minimizes this quantity uniformly over $\nu$. 
Therefore, alternative performance criteria are adopted, most notably Bayesian formulations or minimax approaches that treat $\nu$ as an unknown parameter.

\citet{lorden-ams71} was the first to address the minimax change detection problem. He considered the class of procedures
$\class_\gamma = \{T : \ARLFA(T) \ge \gamma\}$ with $\gamma \ge 1$,
and measured detection delay by the worst‐case (double‐supremum) delay
\begin{equation}\label{eq:SADD-Lorden-def}
\ESADD(T) 
= \sup_{0 \le \nu < \infty}
  \esssup_{X_1,\dots,X_{\nu-1}}
  \Eb_\nu\bigl[(T - \nu + 1)^+ \mid X_1,\dots,X_{\nu-1}\bigr].
\end{equation}
His minimax criterion is
$
\inf_{T:\,\ARLFA(T)\ge\gamma} \ESADD(T).
$

Lorden showed that CUSUM $T_{a_\gamma}$ with $a_\gamma=\log\gamma$ is in $\class_\gamma$ and 
\[
\inf_{T\in\class_\gamma}\ESADD(T)\sim \frac{\log\gamma}{I}\sim\ESADD(T_{a_\gamma}),\quad \gamma\to\infty,
\]
where $I=\Eb_1[Z_1]$.  
\citet{MoustakidesAS86} proved its strict minimax optimality for all $\gamma\ge1$ with $\ARLFA(T_{a_\gamma})=\gamma$.  
\citet{PollakAS85} proposed instead minimizing 
\[
\SADD(T)=\sup_{\nu\ge1}\Eb_\nu[T-\nu+1\mid T\ge\nu] \quad \text{ over $T \in \class_\gamma$}.
\]

An alternative is the Shiryaev--Roberts procedure
\[
T^*_A=\inf\{n\ge1:R_n\ge A\},\quad 
R_n=(1+R_{n-1})e^{Z_n},\;R_0=0.
\]

As shown in \citet{tartakovskypolpolunch-tpa11}, the Shiryaev--Roberts (SR) procedure is asymptotically second-order optimal with respect to Pollak's \(\SADD(T)\) criterion; that is,
\[
\inf_{T \in \class_\gamma} \SADD(T) = \SADD(T^*_{A_\gamma}) + O(1) \quad \text{as } \gamma \to \infty.
\]

\citet{PollakAS85} modified the SR procedure by initializing $R_0$ from its quasi‐stationary distribution under $\Hyp_\infty$. The resulting SRP procedure $T^P_A$ is third‐order asymptotically minimax optimal:
\[
\inf_{T\in\class_\gamma}\SADD(T)=\SADD(T^P_{A_\gamma})+o(1),\quad \gamma\to\infty.
\]
Similarly, \citet{tartakovskypolpolunch-tpa11} proved third‐order optimality for the SR–$r$ procedure with $R_0=r$, where $r$ is a specially designed fixed point.

\citet{LaiIEEE98} extended the Lorden--Moustakides theory to general non-i.i.d.\ models with dependence, where with pre- and post-change conditional densities $f_{0,t}(X_t\mid \Xb^{t-1})$ and $f_{1,t}(X_t\mid \Xb^{t-1})$ the CUSUM statistic is
\[
W_n = \max_{1 \le \nu \le n} \lambda_n^\nu, 
\qquad 
\lambda_n^\nu = \sum_{t=\nu}^n \log \frac{f_{1,t}(X_t\mid \Xb^{t-1})}{f_{0,t}(X_t\mid \Xb^{t-1})}.
\]
Using change-of-measure and the strong law for LLRs, Lai showed that if $n^{-1}\lambda_n^\nu \to I > 0$ a.s.\ and regularity conditions hold, the CUSUM procedure remains first-order asymptotically minimax optimal in the non-i.i.d.\ case \cite[Thms.~1,4(ii)]{LaiIEEE98}.

The following theorem provides necessary details.

\begin{theorem}[CUSUM Asymptotic Optimality, Non-i.i.d.]\label{Th:CPDoptCUSUM}
Suppose 
\begin{equation}\label{Cond1}
\frac{1}{t}\lambda_{\nu+t}^\nu \xrightarrow[t\to\infty]{\text{\rm in $\Pb_\nu$-probability}} I > 0.
\end{equation}
If the right-tail condition
\[
\lim_{n\to\infty} \sup_{\nu \ge 1} \esssup \Pb_\nu \Bigl\{ \max_{1\le t\le n} \frac{1}{n} \lambda_{\nu+t}^\nu \ge I(1+\varepsilon) \,\big|\, \Xb^{\nu-1} \Bigr\} = 0
\]
holds, then as $\gamma\to\infty$,
\[
\inf_{T \in \class_\gamma} \ESADD(T) \ge \frac{\log \gamma}{I}(1+o(1)).
\]

If, in addition, the left-tail condition
\begin{equation}\label{Cond2}
\lim_{n\to\infty} \sup_{1\le \nu\le k} \esssup \Pb_\nu \Bigl\{ \frac{1}{n} \lambda_{k+n}^k \le I-\varepsilon \,\big|\, \Xb^{k-1} \Bigr\} = 0
\end{equation}
holds, then the CUSUM $T_{a_\gamma}$ with $a_\gamma = \log \gamma$ attains this bound:
\[
\ESADD(T_{a_\gamma}) \sim \frac{\log \gamma}{I} \quad \text{as } \gamma \to \infty.
\]
\end{theorem}

The same conclusion, of course, holds for Pollak's less pessimistic maximal expected detection delay measure, \( \SADD(T) \), as well as for the Shiryaev–Roberts procedure.

As mentioned above, the conditional detection delay 
$\Eb_\nu[T - \nu + 1 \mid T \ge \nu]$ cannot be minimized simultaneously for all $\nu$ 
under the constraint $\ARLFA(T) \ge \gamma$. 
\citet{LaiIEEE98} proposed an alternative criterion, based on the maximal unconditional 
false alarm probability,
$
\sup_{k \ge 1} \Pb_\infty(k \le T < k+m),
$
and replacing the conditional delay with the unconditional expected detection delay,
$
\EDD_\nu(T) = \Eb_\nu[(T-\nu+1)^+].
$

Lai further proposed setting the time window \( m = m_\alpha \) depending on the false alarm probability constraint \( \alpha \), such that
\begin{equation}\label{malpha}
\liminf_{\alpha \to 0} \frac{m_\alpha}{|\log \alpha|} \ge \frac{1}{I} 
\quad \text{and} \quad 
\lim_{\alpha\to0}\frac{\log m_\alpha}{|\log \alpha|} = 0.
\end{equation}
Then, the corresponding class of detection procedures is defined as
\[
\class_\alpha^{m_\alpha} = \class(\alpha) = \left\{T : \sup_{k \ge 1} \Pb_\infty(k \le T < k + m_\alpha) \le \alpha \right\}, \quad 0 < \alpha < 1.
\]

He then showed that if, instead of the essential supremum condition \eqref{Cond1}, the following condition holds:
\begin{equation}\label{Cond1noeessup}
\lim_{n \to \infty} \sup_{\nu \ge 1} \Pb_\nu \left\{ \max_{1 \le t \le n} \frac{1}{n} \lambda_{\nu+t}^\nu \ge I(1 + \varepsilon) \right\} = 0 \quad \text{for all} ~ \varepsilon >0,
\end{equation}
then for any \( T \in \class(\alpha) \), as \( \alpha \to 0 \),
\begin{equation}\label{CPDLBuniform}
\EDD_\nu(T) \ge \left \{\frac{\Pb_\infty(T \ge \nu)}{I} + o(1)\right\} |\log \alpha| \quad \text{uniformly in}~ \nu \ge 1.
\end{equation}
See Theorem 2 in \citet{LaiIEEE98}. 

Note that condition~\eqref{Cond1noeessup} holds if $n^{-1} \lambda_{\nu+n}^\nu \to I$ $\Pb_\nu$-a.s.  

In the non-i.i.d.\ case, the CUSUM recursion generally fails, making computation costly. \citet{LaiIEEE98} proposed a window-limited CUSUM (WLCUSUM) over a sliding window of size $m_\alpha$:
\[
W_n^m = \max_{n-m_\alpha \le \nu \le n} \sum_{t=\nu}^{n} \log \frac{f_{1,t}(X_t \mid \Xb^{t-1})}{f_{0,t}(X_t \mid \Xb^{t-1})}, 
\quad
T_a^m = \inf\{ n \ge m_\alpha : W_n^m \ge a \}.
\]
\ignore{
Its local maximal false alarm probability satisfies
\[
\sup_{k \ge 1} \Pb_\infty(k \le T_a^m \le k+m_\alpha) \le 2 m_\alpha e^{-a}.
\]
}
Theorem~4 in \citet{LaiIEEE98} shows that $T_a^m$ is first-order asymptotically optimal as $\alpha \to 0$, attaining the lower bound~\eqref{CPDLBuniform} when $a = a_\alpha = \log(2 m_\alpha / \alpha)$ and the left-tail condition~\eqref{Cond2} holds.
 
In practical applications, the pre-change conditional densities \( f_{0,n}(X_n \mid \Xb^{n-1}) \) are often known, while the post-change densities \( f_{1,n}(X_n \mid \Xb^{n-1}) \) are rarely known—they typically depend on unknown parameters \( \theta \). Let \( \{ f_{\theta,n}(X_n \mid \Xb^{n-1}) , \theta \in \Theta \} \) be a parametric family of conditional densities, where the pre-change parameter \( \theta_0 \in \Theta \) is known and the post-change parameter \( \theta \in \Theta \) is unknown. Then we have
\[
f_{0,n}(X_n \mid \Xb^{n-1}) = f_{{\theta_0,n}}(X_n \mid \Xb^{n-1}), 
\quad 
f_{1,n}(X_n \mid \Xb^{n-1}) = f_{\theta,n}(X_n \mid \Xb^{n-1}).
\]

Let $\Pb_{\nu,\theta}$ and $\Eb_{\nu,\theta}$ denote probability and expectation when the change occurs at $\nu$ with post-change parameter $\theta$, and let $\EDD_{\nu,\theta}(T) = \Eb_{\nu,\theta}[(T-\nu+1)^+]$. Define the log-likelihood ratio
\[
\lambda_n^\nu(\theta) = \sum_{t=\nu}^n \log \frac{f_{\theta,t}(X_t \mid \Xb^{t-1})}{f_{{\theta_0,t}}(X_t \mid \Xb^{t-1})},
\]
assuming $n^{-1} \lambda_{\nu+n}^\nu(\theta) \to I_\theta > 0$ a.s.\ under $\Pb_{\nu,\theta}$.

For a prior $\pi(\theta)$, the mixture likelihood ratio is
\[
\overline{\Lambda}_n^\nu = \frac{\int_\Theta \prod_{t=\nu}^n f_{\theta,t}(X_t \mid \Xb^{t-1}) \, \drm \pi(\theta)}{\prod_{t=\nu}^n f_{{\theta_0,t}}(X_t \mid \Xb^{t-1})}, \quad \bar{\lambda}_n^\nu = \log \overline{\Lambda}_n^\nu.
\]
The window-limited mixture CUSUM is
\[
\overline{T}_a = \inf\left\{ n \ge m_\alpha : \max_{n-m_\alpha \le \nu \le n} \bar{\lambda}_n^\nu \ge a \right\}, \quad \inf \varnothing = \infty.
\]

By Doob's inequality, $\sup_{k\ge1} \Pb_\infty(k \le \overline{T}_a < k+m_\alpha) \le 2 m_\alpha e^{-a}$, so $a = \log(2 m_\alpha/\alpha)$ ensures $\overline{T}_a \in \class(\alpha)$.

\citet[Th.~5]{LaiIEEE98} showed that if $m_\alpha \to \infty$ and $\log m_\alpha / |\log \alpha| \to 0$ as $\alpha \to 0$, and if for each $\varepsilon>0$  there exist a measurable subset \( \Theta_\varepsilon \subseteq \Theta \) and an integer \( n_\varepsilon \) such that \( \pi(\Theta_\varepsilon) > 0 \) and the following condition holds:
\[
\sup_{n \ge n_\varepsilon} \sup_{1 \le \nu \le k} \esssup \Pb_{\nu,\theta} \left\{ \frac{1}{n} \inf_{\theta \in \Theta_\varepsilon} \lambda_{k+n}^k(\theta) < I_\theta - \varepsilon \,\middle|\, \Xb^{k-1} \right\} \le \varepsilon,
\]
then
\[
\EDD_{\nu,\theta}(\overline{T}_{a_\alpha}) \le \frac{\Pb_\infty(\overline{T}_{a_\alpha} \ge \nu)}{I_\theta} |\log \alpha| (1 + o(1)) \quad \text{uniformly in }\nu.
\]
Thus,  by the lower bound \eqref{CPDLBuniform}, the WLCUSUM $\overline{T}_{a_\alpha}$ is asymptotically optimal.

\subsection{Lai's Changepoint Detection--Isolation Theory for Non--i.i.d.\ Data}\label{ssec:CPDISnoniid}

We consider the quickest changepoint detection problem in the \emph{multidecision detection--isolation} (or \emph{classification--identification}) setting with $N$ possible post-change hypotheses. The goal is to design procedures that asymptotically optimize detection delay while controlling false alarms and misidentifications \citep{NikiforovIEEEIT95, Tartakovsky_book2020, TNB_book2014}.

Let $X_1, X_2, \dots$ be observations with change at $\nu$, and let $\Hyp_i$, $i=1,\dots,N$, denote post-change hypotheses. Denote by $\Pb_\nu^i$ and $\Eb_\nu^i$ the measure and expectation under $\Hyp_i$, and by $\Pb_\infty$, $\Eb_\infty$ the no-change scenario.

A sequential change detection--isolation rule $\delta = (T,d)$ consists of a stopping time $T$ and a terminal decision $d \in \{1,\dots,N\}$, where $d=i$ indicates $\Hyp_i$ is accepted at $T$. In the i.i.d.\ case, observations are independent with pre-change density $f_0$ and post-change density $f_i$ under $\Hyp_i$.

In the i.i.d.\ case, \citet{NikiforovIEEEIT95} introduced the first minimax theory for change detection and isolation, using the ARL to false alarm or false isolation as risk measures. Observations follow $\Pb_1^i$ if $\nu=1$ and $\Hyp_i$ is true, and follow $\Pb_\infty$ under the nominal regime.

Consider the following sequence of alarm times and final decisions \( (T_r, d_r) \):
\[
T_0 = 0 < T_1 < T_2 < \cdots < T_r < \cdots, \quad\text{and}\quad d_1,\, d_2,\, \ldots,\, d_r,\, \ldots,
\]
where \( T_r \) is the alarm time of the detection–isolation algorithm applied to the sequence \( X_{T_{r-1}+1}, X_{T_{r-1}+2}, \ldots \).

Let \(X_1, X_2, \dots\) be i.i.d.\ with pre-change density \(f_0\) and post-change density \(f_i\) under hypothesis \(\Hyp_i\), \(i=1,\dots,N\). The ARL to the first false alarm of type \(j\) is \(\Eb_\infty[\inf\{T_r: d_r=j\}]\), and the ARL to false isolation of type \(j\ne \ell\) under \(\Hyp_\ell\) is \(\Eb_1^\ell[\inf\{T_r: d_r=j\}]\). Define the class
\begin{equation}\label{Nik1}
\class_\gamma = \Bigl\{ \delta=(T,d) : \min_{0\le \ell \le N} \min_{j\ne \ell} \Eb_1^\ell[\inf\{T_r:d_r=j\}] \ge \gamma \Bigr\},
\end{equation}
where $\Eb_1^0=\Eb_\infty$. The worst-case expected detection delay is
\[
\ESADD(\delta) = \max_{1\le i \le N}\sup_{\nu\ge 1} \esssup \Eb_\nu^i[(T-\nu+1)^+ \mid X_1,\dots,X_{\nu-1}],
\]
and the minimax goal is \(\min_{\delta\in\class_\gamma} \ESADD(\delta)\), usually in the asymptotic regime \(\gamma \to \infty\).  

\citet{NikiforovIEEEIT95} proposed a multi-hypothesis GLR test that is first-order asymptotically optimal but computationally expensive. Later, \citet{NikiforovIEEEIT00, NikiforovIEEEIT03} introduced a minimax formulation using the maximal conditional false isolation probability,
$
\sup_{\nu\ge1} \Pb_\nu^\ell(d=j\ne \ell \mid T>\nu),
$
with ARL to false alarm \(\ARLFA(T) = \Eb_\infty[T]\), leading to the class
\[
\class_{\gamma,\alpha} = \Bigl\{ \delta : \ARLFA(T) \ge \gamma,\; \max_{i\ne j} \sup_{\nu\ge 1} \Pb_\nu^i(d=j \mid T>\nu) \le \alpha \Bigr\}.
\]
He proposed a matrix CUSUM procedure minimizing \(\max_i \SADD_i(T) = \max_i \sup_\nu \Eb_\nu^i[T-\nu+1 \mid T \ge \nu]\) asymptotically as \(\gamma \to \infty, \alpha \to 0\). \citet{Tartakovsky-SQA08b} extended this to per-hypothesis constraints \(\alpha_i\), with an efficient matrix-based procedure minimizing \(\SADD_i(\delta)\) for all \(i\) as \(\gamma \to \infty\) and \(\max_i \alpha_i \to 0\).  

We now turn to the non-i.i.d.\ setting and discuss Lai's important contributions to the theory of change detection and isolation in this more general framework. Assume that the observations are dependent and non-identically distributed, with the pre-change conditional density \( f_{0,t}(X_t \mid \Xb^{t-1}) \) for \( t < \nu \), and the post-change conditional density 
\( f_{i,t}(X_t \mid \Xb^{t-1}) \) for \( t \ge \nu \), when hypothesis \( \Hyp_i \) is correct (\( i = 1, \ldots, N \)).

\citet{LaiIEEE00} extended \citet{NikiforovIEEEIT95} by imposing maximal local constraints for all \(\nu \ge 1\).  
For \(i \in \{0,\dots,N\}, j \in \{1,\dots,N\}\), define
\[
\lambda_n^\nu(i,j) = \sum_{t=\nu}^n \log \frac{f_{i,t}(X_t\mid\Xb^{t-1})}{f_{j,t}(X_t\mid\Xb^{t-1})}.
\]

The class controlling local false alarm/isolation is
\begin{align*}
\class_\alpha(m_\alpha) = \Bigl\{ \delta : &~
\sup_{k\ge 1} \Pb_\infty(k \le T < k+m_\alpha) \le \alpha m_\alpha, \\[0.5ex]
&~ \max_{1 \le i \le N} \sup_{\nu\ge 1} 
\Pb_\nu^i(\nu \le T < \nu+m_\alpha,\, d \ne i) \le \alpha m_\alpha 
\Bigr\}.
\end{align*}

The procedure \(\delta_a(m_\alpha) = (T_a, d_a)\) has
\begin{align*}
T_a & = \inf \Bigl\{ n : \exists 1 \le i \le N,~ \max_{n-m_\alpha \le k \le n} \lambda_n^k(i,0) - 
\bigl(\max_{\ell\ne i} \max_{n-m_\alpha \le k \le n} \lambda_n^k(\ell,0)\bigr)^+ \ge a \Bigr\},
\\
d_a & = \argmax_{1 \le i \le N} \max_{T_a-m_\alpha \le k \le T_a} \lambda_{T_a}^k(i,0).
\end{align*}

If $m_\alpha = O(|\log \alpha|)$, $a = \log(2N/\alpha)$, and certain regularity conditions hold for the log-likelihood ratios, then the procedure $\delta_a(m_\alpha)$ is uniformly asymptotically optimal in the class $\class_\alpha(m_\alpha)$. 
For details, see Theorem~7 in \citet{LaiIEEE00}.

\section{MULTIPLE CHANGE-POINT SURVEILLANCE AND ESTIMATION}\label{sec:MCPS}

In this section, we review Lai's contributions to 
Bayesian analysis of multiple change-points.

\subsection{An Extended Shiryaev's Rule for Composite Hypotheses}

Suppose that the observations $\{ X_n \}_{n\ge 1}$ are 
independent and such that $X_1, \dots, X_{\nu-1}$ are 
each distributed according to a common density $f_0(x)$,
while $X_{\nu}, X_{\nu+1}, \dots$ each follows a 
common density $f_1(x) \neq f_0(x)$. Moreover, assume 
that the change point $\nu$ has a geometric prior 
distribution with success probability $p$, that is, 
$\Pb(\nu=k) = p(1-p)^{k-1}$ for $k=1,2,\dots$.

We first consider the case of two simple hypotheses, where $f_1$ and $f_0$ are fully known. 
Assigning a loss of $c$ for each observation taken after the change point $\nu$ and a loss of $1$ 
for a false alarm before $\nu$, \citet{shiryaev-tpa1963, Shiryaev1978} showed, using optimal 
stopping theory, that the optimal detection procedure compares the posterior probability 
of a change with a threshold. 

This leads to the {\it Shiryaev procedure}, which has the form
\begin{equation}\label{shiryaev.rule.equ3}
T_{\rm S}(\eta) = \inf \Bigl\{ n \ge 1 : R_{n,p} \ge \eta \Bigr\}, \qquad
R_{n,p} = \sum_{k=1}^n \prod_{i=k}^n \frac{f_1(X_i)}{(1-p)\, f_0(X_i)}.
\end{equation}

\noindent where $\eta$ is chosen in such a way that 
the {\it probability of false alarm} (PFA) is
exactly equal to $\alpha$. For large values of the threshold 
$\eta$, the average delay to detection of
the Shiryaev procedure can be approximated 
to first order as $\eta \rightarrow \infty$; 
see \citet[Theorem 7.1.4]{TNB_book2014}.
Moreover, as the geometric prior parameter vanishes ($p \to 0$), 
the Shiryaev 
statistic converges to the {\it Shiryaev–Roberts statistic}, and 
the corresponding {\it Shiryaev–Roberts rule} is asymptotically 
Bayes risk efficient \citep{PollakAS85}.

Note that the Shiryaev procedure extends naturally to dependent data (see, e.g., \cite{TarIEEE2017, TarVeerIEEE2005}).

We now consider the case of two composite hypotheses,
that is, $f_1$ and $f_0$ are only partially known. 
Suppose that $f_1$ and $f_0$ are not known in advance, 
but belong to a multivariate exponential family
\begin{equation}\label{dist.exp.family}
f_{\theta}(X) = \exp \{ \theta' X - \psi (\theta) \}.
\end{equation}
 
\noindent Let $\pi$ be a prior density function 
\begin{equation}\label{dist.exp.family.prior}
\pi(\theta; a_0, \mu_0) = c(a_0, \bmu_0)\,
\exp\big\{ a_0 \mu_0' \theta - a_0 \psi(\theta) \big\},
\qquad \theta \in \Theta,
\end{equation}

\noindent in which $\Theta$ is the parameter space and 
\[
\frac{1}{c(a_0, \bmu_0)} = \int_{\Theta} 
\exp \big\{ a_0 \mu_0' \theta - a_0 \psi(\theta) \big\}\, \drm\theta,
\qquad \bmu_0 \in (\nabla \psi)(\Theta),
\]

\noindent and $\nabla$ denotes the gradient vector of partial 
derivatives.  The posterior density of $\theta$ given 
the observations $X_1, \dots, X_m$ drawn from
$f_{\theta}$ is
\begin{equation}\label{dist.exp.family.post}
\pi \!\left( \theta;\, a_0+m,\,
\frac{a_0 \mu_0 + \sum_{i=1}^m X_i }{a_0 + m} \right).
\end{equation}

\noindent Therefore, \eqref{dist.exp.family.prior} is a 
conjugate family of priors and the predictive distribution 
satisfies
\[
\int_{\Theta} f_{\theta}(X)\, \pi(\theta; a, \mu)\, \drm\theta 
= \frac{c(a, \mu)}{c\!\left(a+1, \tfrac{a \mu + X}{a+1}\right)}.
\]

Suppose that the parameter $\theta$ takes the value $\theta_0$
for $t<\nu$ and another value $\theta_1$ for $t \ge \nu$, and that 
the change-time $\nu$ and the pre- and post-change values $\theta_0$
and $\theta_1$ are unknown. Following 
\cite{shiryaev-tpa1963, Shiryaev1978}, one can use the Bayesian 
approach that assumes $\nu$ to be geometric with parameter $p$
but constrained to be larger than $n_0$ and that $\theta_0, 
\theta_1$ are independent, have the same density function \eqref{dist.exp.family.prior} and are also independent
of $\nu$. 

Let $\pi_n = \Pb\{ \nu \le n \mid X_1, \dots, X_n \}$. 
Whereas $\pi_n$ is a Markov chain in the case of known parameters 
$\theta_0$ and $\theta_1$, it is no longer Markovian in the present setting with
unknown pre- and post-change parameters. Consequently, Shiryaev’s rule,
which triggers an alarm once $\pi_n$ exceeds a threshold, is no longer optimal; see 
\citet[pp.~540--541]{Zacks1991}, who suggests applying dynamic programming to 
determine the optimal stopping rule, but also notes that ``it is generally 
difficult to determine the optimal stopping boundaries.'' 
Due to this complexity, \citet{ZacksBarzily1981} 
introduced a more tractable myopic (two-step-ahead) policy in the 
univariate Bernoulli case ($X_i \in \{0,1\}$).

\cite{LaiXing2010} introduced a modification of Shiryaev's 
rule and showed it is asymptotically Bayes as 
$p\rightarrow 0$. Let ${\cal F}_t$ be the $\sigma$-field 
generated by $X_1, \dots, X_t$ and 
$$
\pi_{0,0}=c(a_0, \mu_0), \quad \pi_{i,j} = c
\Big(a_0 + j-i+1, \frac{a_0 \mu_0 + \sum_{t=i}^j 
X_t}{a_0+j-i+1} \Big). 
$$

\noindent Note that for $n_0 < i \le n$,
\begin{equation}\label{exShi.equ1}
\Pb\{ \nu = i |{\cal F}_n\} \propto p(1-p)^{i-1} \pi_{0,0}^2 
\big/ \pi_{1,i-1} \pi_{i,n}, \quad
\Pb\{ \nu >n |{\cal F}_n\} \propto (1-p)^n \pi_{0,0} \big/ \pi_{1,n}.
\end{equation}

\noindent The normalizing constant is determined by the fact 
that all the probabilities in \eqref{exShi.equ1} sum to 1.
Let $p_{i,n}=\Pb\{ \nu=i| {\cal F}_n \}$ be the posterior
probability given the observed samples up to time $n$, 
we then have
$$
\Pb(n_0 < \nu \le n | {\cal F}_n) = \sum_{i=n_0+1}^n p_{i,n}
= \frac{ \sum_{i=n_0+1}^n \Pb(\nu=i | {\cal F}_n) }{
\sum_{i=n_0+1}^n \Pb(\nu=i| {\cal F}_n)
+ \Pb(\nu>n | {\cal F}_n) },
$$ 

\noindent in which $\Pb\{ \nu = i |{\cal F}_n\}$ and 
$\Pb\{ \nu >n |{\cal F}_n\}$ are given by \eqref{exShi.equ1}.
Therefore, Shiryaev's stopping rule in the present setting 
of unknown pre- and post- change parameters can again 
be written in the form of \eqref{shiryaev.rule.equ3}
with 
$$
R_{n,p}=\sum_{i=n_0+1}^n \frac{\pi_{0,0} \pi_{1,n}}{
(1-p)^{n-i} \pi_{1,i}\pi_{i,n} }.
$$ 

\noindent \cite{LaiXing2010} proposed an 
extended Shiryaev's rule
\begin{equation}\label{exShiryaev.rule1}
T_{\rm exShi}=\inf \{ n>n_p: \Pb(\nu\le n| \nu\ge n-k_p, {\cal F}_n) \ge \eta_p \},
\end{equation}

\noindent and showed that it is asymptotically optimal 
as $p\rightarrow 0$, for suitably chosen $k_p$, $\eta_p$ 
and $n_p \ge n_0$; see \citet[Theorem 3.3]{LaiXing2010}. 
Since
$$
\Pb(\nu\le n | \nu\ge n-k_p, {\cal F}_n) = \frac{ 
\sum_{i=n-k_p}^n \Pb(\nu=i | {\cal F}_n) }{ \sum_{i=n-k_p}^n 
\Pb(\nu=i| {\cal F}_n) + \Pb(\nu>n | {\cal F}_n)},
$$

\noindent we can use \eqref{exShi.equ1} to rewrite \eqref{exShiryaev.rule1} in the form 
\begin{equation}\label{exShiryaev.rule2}
T_{\rm exShi}=\inf \Big\{ n > n_p: \sum_{i=n-k_p}^n \frac{\pi_{0,0}
 \pi_{1,n}}{(1-p)^{n-i+1}
\pi_{1,i-1}\pi_{i,n}} \ge \gamma_p \Big\}. 
\end{equation}

\noindent This has essentially the same form as Shiryaev's rule 
\eqref{shiryaev.rule.equ3} with the
obvious changes to accommodate the unknown $f_{\theta_0}$ and 
$f_{\theta_1}$, and with the important sliding window modification
$\sum_{i=n-k_p}^n$ of Shiryaev's sum $\sum_{i=n_0+1}^n$, which has too 
many summands to trigger false alarms when $\theta_0$ and $\theta_1$ 
are estimated sequentially from the observations.

\subsection{Surveillance of Multiple Change-Points}

Quickest detection and sequential surveillance are 
closely related but differ in scope and objectives. 
Quickest detection targets a single change in a process 
and is widely applied in areas such as fault detection, 
navigation integrity monitoring, and radar/sonar signal 
processing. Sequential surveillance, by contrast, addresses 
multiple changes over time---commonly called multiple 
change-point detection---with applications in finance, 
cybersecurity, and public health. A common strategy is 
detection–segmentation, where each detected change-point 
becomes the starting point for the next search. However, 
false alarms or detection delays at one stage can propagate 
and degrade the performance of subsequent detections.

\cite{LaiLiuXing2009} introduced a Bayesian model for 
multiple change-points in multivariate exponential 
family distributions with unknown pre- and post-change 
parameters---a key setting in sequential surveillance. 
Assume observations $X_1,\dots,X_t,\dots$ follow the multivariate exponential family \eqref{dist.exp.family}
where $\theta_t$ may change occasionally, with indicators $I_t 
= 1_{{\theta_t \neq \theta_{t-1}}}$ that are i.i.d. 
Bernoulli$(p)$. When $I_t=1$, the new $\theta_t$ is 
sampled from $\pi$. The conjugacy of the prior yields 
explicit formulas for both sequential (filtering) 
estimates $\Eb(\mu_t|{\bf X}^t)$ and fixed-sample 
(smoothing) estimates $\Eb(\mu_t|{\bf X}^n)$, where 
$\mu_t=\nabla\psi(\theta_t)$.

For the above model, \citet{LaiXing2008b, LaiXing2011} 
proposed a general hidden Markov filtering approach that 
yields recursive and tractable estimators of multiple 
change-points and is computationally very efficient. 
Let $K_t = \max \{ s\le t: I_s =1 \}$ be the most 
recent change-time $K_t$ up to $t$. Denote by $p_{it} = 
P(K_t = i |{\cal Y}_t)$ the probability that the most recent 
change-time up to time $t$ is $K_t$ and $f(\cdot | \cdot)$
the conditional density. Since $K_t$ can take values from
1 to $t$, \cite{LaiXing2011} showed that the posterior 
distribution of $\theta_t$ given ${\bf X}^t$ is a mixture 
of distributions,
\begin{equation}\label{mcp.equ6.2}
f(\theta_t |{\bf X}^t) = \sum_{i=1}^t p_{it} \pi(\theta_t; 
a_0+t-i+1, \bar{X}_{i,t} ). 
\end{equation}

\noindent where $\bar{X}_{i,j} = (a_0 \mu_0 + \sum_{k=i}^j 
X_k) \big/ (a_0 + j-i+1)$, $j\ge i$, is the posterior
mean, and the mixture weights $p_{it}$ can be represented 
recursively by 
\begin{equation}\label{mcp.equ7}
p_{it} = \frac{p_{it}^*}{\sum_{j=1}^t p_{jt}^*}, 
\qquad  p_{it}^* = \left\{ \begin{array}{ll} p  \pi_{0,0} / 
\pi_{t,t} & \mbox{if } i=t, \\ (1-p)  p_{i,t-1} 
\pi_{i,t-1} / \pi_{i,t} &
\mbox{if } i<t, \end{array} \right. 
\end{equation}

\noindent in which $\pi_{0,0}=c(a_0, \mu_0)$ and $\pi_{i,j} 
= c(a_0 + j-i+1, \bar{X}_{i,j} )$. Then the change-point 
probability, the probability that the most recent
change-point occurs at one of the times $\{s, s+1, 
\dots, t\}$, and the posterior mean at $t$ are 
given, respectively, by 
\begin{equation}\label{mcp.filter.est}
\Pb(I_t=1|{\bf X}^t) = p_{tt}, \quad
\Pb( s\le K_t \le t) = \sum_{i=s}^t p_{it}, \quad
\Eb(\mu_t | {\bf X}^t) = \sum_{i=1}^t p_{it} 
\bar{X}_{i,j}.
\end{equation}

To obtain a surveillance rule, \cite{LaiLiuXing2009} 
use sliding windows $\sum_{i=n-k_p}^n$ as in 
\eqref{exShiryaev.rule2} but with the summands 
modified to be the posterior probabilities $p_{in}$ 
that the most recent change-point up to time $n$ 
occurs at $i$. Then the surveillance rule based on
the assumption of {\it multiple change-points} (MCP) 
can be expressed as
\begin{equation}\label{mcp.rule2}
T_{\rm MCP}=
\inf \big\{ n> n_p: \sum_{i=n-k_p}^n p_{in} \ge \gamma_p \big\}. 
\end{equation}

\noindent \citet{LaiLiuXing2009} showed that, by suitably choosing 
$k_p$ and $\gamma_p$, the surveillance rule \eqref{mcp.rule2} 
achieves asymptotically optimal Bayes and frequentist performance, 
using the definitions of false alarm rate and detection delay 
for multiple change-points.  

In practice, two main issues arise when implementing the surveillance rule \eqref{mcp.rule2} as $n$ grows: estimating the hyperparameters $p$, $a_0$, and $\mu_0$, and handling computational cost. Since the forward filter $\theta_t|{\bf X}^t$ depends on these hyperparameters, \cite{LaiXing2011} proposed an empirical Bayes approach. From $p_{it}$, the likelihood of $(p,a_0,\mu_0)$ is
\begin{equation}\label{mcp.llh}
\prod_{t=1}^n f(X_t | {\bf X}_{t-1}) = \prod_{t=1}^n 
\Big( \sum_{i=1}^t p_{it}^* \Big),
\end{equation}

\noindent Because $X_t$ are exchangeable with mean $\mu_0$, we estimate $\mu_0$ by the sample mean $\widehat{\mu}=n^{-1}\sum_{t=1}^n X_t$. A simple choice $a_0=1$ treats $\widehat{\mu}$ as one pseudo-observation at a change-point. The key parameter is $p$, the relative frequency of change-points. Substituting $a_0=1$ and $\widehat{\mu}$ into \eqref{mcp.llh}, $p$ can be estimated by maximizing $l(p)=\sum_{t=1}^n \log (\sum_{i=1}^t 
p_{it}^*)$, using a grid search over ${2^j/n: j_0 \le j \le j_1}$.

To reduce the linear computational complexity in 
\eqref{mcp.equ7}, \cite{LaiXing2011} proposed the 
{\it bounded complexity mixture} (BCMIX) approximation. 
The idea is to retain the most recent $m$ weights, discard 
the smallest among the remaining ones, and then reweight 
at each step. Let ${\cal K}_{t-1}$ be the set of indices 
kept at stage $t-1$ (with ${\cal K}_{t-1} \supset {t-1, 
\dots,t-m}$). At stage $t$, compute $p_{i,t}$ as in 
\eqref{mcp.equ7} for $i \in {t} \cup {\cal K}_{t-1}$, 
identify the smallest $p_{i,t}$ among indices $\le t-m$, 
remove it, and define
$$
{\cal K}_t = \{ t \} \cup ({\cal 
K}_{t-1} - \{ i_t \}), \quad
p_{i,t} = \Big( p_{i,t}^* \Big/ \sum_{j \in {\cal K}_t } 
p_{j,t}^* \Big), \quad i \in {\cal K}_t.
$$

\noindent Thus, $|{\cal K}_t| \le M$, ensuring bounded complexity regardless of $n$.

The surveillance rule \eqref{mcp.rule2} with BCMIX can monitor complex systems undergoing multiple changes, with applications in finance and beyond. For example, \cite{XingEtAl2020} applied it to detect shifts in firms’ credit rating migration generators.

\subsection{Inference on Multiple Change Points}

The hidden Markov filtering approach underlying the 
surveillance rule \eqref{mcp.rule2} can also be extended 
to estimate $\theta_t$ for each $t=1,\dots,n$ from 
observations $X_1,\dots,X_n$. This yields the smoothing 
estimate of $\theta_t$ given ${\bf X}^t$. Because the 
number and locations of change-points are unknown, direct 
estimation is computationally demanding. By combining the 
forward and backward filters, however, smoothing estimates 
can be obtained efficiently.

In particular, \cite{LaiXing2011} showed how to derive
the posterior distribution of $\theta_t \mid {\bf X}^n$ 
by applying Bayes’ theorem to combine the forward filter
$\theta_t \mid {\bf X}^t$ with the backward
filter $\theta_t \mid {\bf X}^{t+1,n}$.
The backward filter is obtained by reversing time
and is expressed as 
\begin{equation}\label{mcp.equ8}
f(\theta_t | {\cal Y}_{t+1,n}) = p \pi(\theta_t; a_0, 
\mu_0) + (1-p) \sum_{j=t+1}^n q_{j,t+1}
\pi (\theta_t ; a_0 + j-t, \bar{X}_{t+1,j}), 
\end{equation}

\noindent where $q_{jt} = q_{jt}^*/ \big( \sum_{l=t}^n q_{lt}^*)$ and 
\begin{equation}\label{mcp.equ9}
q_{j,t}^* =\left\{ \begin{array}{ll}
p \pi_{0,0} / \pi_{t,t} & \mbox{if } j=t,\\
(1-p) q_{j,t+1} \pi_{t+1,j}/ \pi_{t, j} & \mbox{if } j>t.
\end{array} \right. 
\end{equation}

\noindent By Bayes' theorem, the smoother is proportional to 
the product of forward and backward filters divided 
by the prior distribution, that is, 
\begin{equation}\label{mcp.bayes}
f(\theta_t | {\bf X}^n) \propto f(\theta_t | {\bf X}^t) 
f(\theta_t | {\bf X}^{t+1,n})  \big/ \pi(\theta; a_0, \mu_0).
\end{equation}

\noindent Combining \eqref{mcp.equ8} with \eqref{mcp.equ6.2}, and noting that 
$$
\pi \big(\theta; a_0+t-i+1, \bar{X}_{i,t} \big)
\frac{\pi \big(\theta; a_0+j-t, \bar{X}_{t+1,j} \big) }{
\pi \big(\theta; a_0, \mu_0 \big) }
= \frac{\pi_{it} \pi_{t+1,j}}{\pi_{ij} \pi_{00} }
\pi \big(\theta; a_0+j-i+1, \bar{X}_{ij} \big),
$$

\noindent one can use \eqref{mcp.bayes} to obtain the smoother 
$\theta_t | {\bf X}^n$, which is expressed as
\begin{equation}\label{mcp.sm1}
f(\theta_t | {\bf X}^n) = \sum_{1\le i \le t \le j \le n}
\beta_{ijt} \pi(\theta_t; a_0 + j-i+1, \bar{X}_{i,j}), 
\end{equation}

\noindent where $\beta_{ijt}= \beta_{ijt}^* \big/ P^*_t$, $P^*_t = p + 
\sum_{1 \le i \le t < j \le n} \beta_{ijt}^*$, and
\begin{equation}\label{mcp.sm2}
\beta_{ijt}^* = \left\{ \begin{array}{ll} 
 p p_{it} & \mbox{if } i\le t=j, \\
(1-p) p_{it} q_{j,t+1} \pi_{it} \pi_{t+1,j} \big/ \pi_{ij} \pi_{00} 
& \mbox{if } i\le t <j.
\end{array} \right. 
\end{equation}

\noindent From the above, it follows that the change-point
probability and posterior mean at time $t$ are expressed
as 
\begin{equation}\label{mcp.sm3}
\Pb(I_{t+1}=1|{\bf X}^n) = p \big/ P^*_t, \qquad
\Eb(\mu_t | {\bf X}^n) = \sum_{1\le i \le t \le j \le n}
\beta_{ijt} \bar{X}_{i,j}. 
\end{equation}

Computing the smoothing estimates using \eqref{mcp.sm3}
for all $t=1, \dots, n$ results in a computational 
complexity $O(n^3)$. \cite{LaiXing2011} showed how to
apply the BCMIX idea to the smoother \eqref{mcp.sm1}.
We first define the forward filter $\theta_t \mid {\bf X}^t$
as in the preceding section. 
For the backward filter $\theta_t \mid {\bf X}^{t+1,n}$, let $\widetilde{\cal K}_{t+1}$ denote the set of indices retained at stage $t+1$ (with $\widetilde{\cal K}_{t+1} \supset {t+1, \dots, t+m}$). At stage $t$, compute $q_{j,t}$ as in \eqref{mcp.equ9} for $j \in {t} \cup \widetilde{\cal K}{t+1}$, identify the smallest $q{j,t}$ among indices $\ge t+m$, remove it, and define
$$
\widetilde{\cal K}_t= \{ t \} 
\cup (\widetilde{\cal K}_{t}- \{ j_t \}), \qquad
q_{j,t} = \Big( q_{j,t}^* \Big/ \sum_{j \in 
\widetilde{\cal K}_t } q_{j,t}^* \Big), \quad 
j \in \widetilde{\cal K}_t.
$$

\noindent This gives a BCMIX approximation to the backward filter $\theta_t \mid {\bf X}^{t+1,n}$. Then, for each $t=1,\dots,n-1$, the BCMIX approximation to the smoother \eqref{mcp.sm1} can be obtained by combining the forward and backward BCMIX filters with selected weight indices ${\cal K}_t$ and $\widetilde{\cal K}_{t+1}$ via Bayes’ theorem:
$$
f(\theta_t | {\bf X}^n) \approx \sum_{i\in {\cal K}_t, \ j\in 
\widetilde{\cal K}_{t+1}} \widetilde{\beta}_{ijt} 
\pi(\theta_t; a_0 +j-i +1, \bar{X}_{i,j}), 
$$

\noindent in which 
$$
\widetilde{\beta}_{ijt}=\beta_{ijt}^* / \widetilde{P}^*_t,
\qquad 
\widetilde{P}^*_t=p + \sum_{1\le t \le n, i\in {\cal K}_t, 
j \in \widetilde{\cal K}_{t+1} }\beta_{ijt}^*,
$$ 

\noindent and
$\beta_{ijt}^*$ is given by \eqref{mcp.sm2} for $i\in {\cal K}_t$ and
$j \in \widetilde{\cal K}_{t+1}$. The BCMIX approximation to 
the change-point probability and posterior mean at time $t$ 
are therefore 
$$
\widehat{\Pb}(I_{t+1}=1 | {\bf X}^n) = \frac{p}{P_t^*}, 
\qquad \widehat{\Eb}( \mu_t | {\bf X}^n) = 
\sum_{i\in {\cal K}_t, \ j \in \widetilde{\cal 
K}_{t+1} } \widetilde{\beta}_{ijt} \bar{X}_{i,j}. 
$$

The BCMIX approximation greatly reduces the computational cost of smoothing estimates for $\theta_t$ ($t=1,\dots,n$). In multiple changepoint problems with large $n$ (e.g., $n \sim 10^6$), the full sequence $\{\theta_t\}_{1\le t \le n}$ can be estimated with $O(n)$ complexity. This efficiency demonstrates that the Markov filtering framework, combined with BCMIX, can be effectively applied to scalable change-point estimation in complex systems.

Hidden Markov models for multiple change points, along with 
their smoothing estimates, provide a powerful framework for 
analyzing complex dynamical systems with abrupt shifts. 
Building on this approach, \citet{LaiXing2013} developed 
changepoint autoregressive GARCH models to infer discrete-
time volatility dynamics and simultaneous changes in 
autoregressive coefficients and volatilities from asset 
prices. Similarly, \citet{XingEtAl2012} modeled firms’ 
credit rating transitions as piecewise homogeneous Markov 
chains with unobserved structural breaks. Beyond finance, 
these models have been applied to genomic sequence data, 
including array-based comparative genomic hybridization 
\citep{LaiXingZhang2008}, parent-specific DNA copy number in 
tumors \citep{ChenXingZhang2011}, and chromatin 
immunoprecipitation sequencing \citep{XingEtAl2012b}.

\section{NONLINEAR RENEWAL THEORY}\label{Sec:NRT}

In this section, we summarize Lai’s contributions to nonlinear renewal theory and to multivariate Markov renewal theory.

\subsection{Nonlinear and Markov Nonlinear Renewal Theories}\label{sec1}

Motivated by the study of boundary crossing times and their importance in sequential analysis, 
nonlinear renewal theory was developed in the seminal works \cite{LaiSiegmund1977,LaiSiegmund1979,Woodroofe1976,Woodroofe1977}. 
Comprehensive treatments of the classical approaches are given in the monographs \cite{Woodroofe1982,siegmund-book85,Tartakovsky_book2020}. 
Building on these foundations, \cite{Zhang1988} established several general results that broadened the scope of the theory. 
More recently, the framework has been extended to the multivariate setting by \cite{FuhKao-SIAMFM21}, thereby opening new directions for applications.

Let $X_1, X_2, \ldots$ be i.i.d.\ random variables with common distribution $F$ and finite, positive mean $\mu = \Eb[ X_1], 0 < \mu < \infty;$ and $S_n = \sum_{k=1}^n X_k,~n \geq 1,$ denote the partial sums.  Let $\{Z_n = S_n + \eta_n, n\ge 1\}$ be a
perturbed random walk in the following sense: $S_n$ is a  random walk, $\eta_n$ is
${\cal F}_n$-measurable, where ${\cal F}_n$ is the $\sigma$-algebra generated by $\{S_k, 1 \le k \le n\}$. Let $\eta_n$ be {\it slowly changing}, i.e. 
$\frac{1}{n} \max_{1 \leq k \leq n} \big\vert \eta_k \big\vert \rightarrow 0~in~probability,$
and for every $\epsilon > 0$, there exist $n^*$ and $\delta > 0$ such that for all $n \geq n^*$,
$\Pb\left\{ \max_{1 \leq k \leq n\delta} |\eta_{n+k} - \eta_n| > \epsilon \right\} < \epsilon.$

Let $A =\{A(t;\lambda), \lambda \in \Lambda\}$ be a family of boundary
functions for some index set $\Lambda$. For each $\lambda \in \Lambda$, define
\begin{eqnarray}
 T = T_{\lambda} = \inf\{n \geq 1: Z_n > A(n;\lambda) \},~~~\inf \varnothing
 = \infty. \label{1.2b}  
\end{eqnarray}
It is easy to see that under the positive drift assumption $\mu > 0$, we have $T_\lambda < \infty$ for all $\lambda > 0$ with probability one. In nonlinear renewal theory, one focuses on asymptotic approximations for the distribution of the overshoot and for the expected stopping time $\Eb[T]$ as the boundary tends to infinity.

To illustrate the motivation of investigating stopping times $T_\lambda$ in (\ref{1.2b}),
we consider the following simple example: let $\eta_n = 0$ and $A(n;\lambda)= \lambda n^\alpha$ for $0 \leq \alpha < 1$, then
\begin{eqnarray}\label{1.3a}
T:=T_\lambda = \inf\{n \geq 1: S_n  > \lambda n^\alpha \}.
\end{eqnarray}
Note that (\ref{1.3a}) is a standard formulation in nonlinear renewal theory, developed in \cite{LaiSiegmund1977,LaiSiegmund1979}, which was originally motivated by problems arising in sequential analysis for statistical models. This formulation plays a central role in analyzing boundary crossing probabilities and provides the foundation for deriving asymptotic approximations of stopping times in a wide range of statistical applications.

By combining renewal theory with Wald's identity, a standard approach is to 
investigate the difference between $T_\lambda$ and a stopping time defined by 
crossing linear boundaries with varying drift. Specifically, we define
\begin{equation}\label{1.3b}
   \tau(c,u) = \inf\{n \geq 1 : S_n - nu > c\}, \quad c \geq 0,~ 0 < u \leq \mu,
\end{equation}
and aim to establish the uniform integrability of $ |T_{\lambda} - \tau(c_{\lambda},d_{\lambda})|^p$, $p \geq 1$,
for suitable choices of $c_{\lambda}$ and $d_{\lambda}$. 
Nonlinear renewal theory is then derived directly from the corresponding 
results in the linear case with varying drift, by leveraging uniform integrability 
and the weak convergence of the overshoot.

When $n$ is the first time of $S_n$ crossing the boundary $\lambda n^\alpha$, by Wald's identity, we have
$n \mu = S_n \approx \lambda n^\alpha \Rightarrow n \approx \big(\lambda/\mu \big)^{1/(1-\alpha)}.$
Denote $b_\lambda= \big(\lambda/\mu\big)^{1/(1-\alpha)}$. By using linearlization,
${d( \lambda n^\alpha)}/{d \alpha} = \alpha  \lambda n^{\alpha - 1} \approx \alpha \mu,$
we can approximate the curve boundary $\lambda n^\alpha$ by
$l_\lambda(n)= \mu b_\lambda + \alpha \mu (n - b_\lambda),$ Taylor's expansion at $b_\lambda$.
Therefore, $c = c_\lambda = \mu (1- \alpha) b_\lambda  > 0$ and $d = d_\lambda =  \alpha \mu$ in this simpe case.
Since we need to find upper and lower bounds of the curve boundary for each $n$,
$\alpha$ is in a range of $[0, 1)$. This implies that we consider a random walk with varying drift or a uniform renewal theory. The reader is referred to the above mentioned articles for details.

Next, we present a multivariate nonlinear renewal theory from \cite{FuhKao-SIAMFM21} as
follows:
let $\{(X_n, Y_n),n=1,2,\cdots\}$ be a sequence of i.i.d.\ random
variables in ${\bf R}^{d+1}$, where $X_1 \in {\bf R^1}$ with positive drift and
finite variance as before, and $Y_1 \in {\bf R}^d$ with $\Eb[Y_1]=\vec{0}$ and
$\mathit{Var}(Y_1) =\Sigma_Y$. Denote $\Sigma$ as the variance--covariance matrix
of $(X_1, Y_1)$. Let  $S_n = \sum_{k=1}^n X_k$, and ${\bm W}_n = \sum_{k=1}^n Y_k$ with ${\bm W}_0=\vec{0}$. 

Consider the stopping time
\begin{equation}\label{tau_general}
\tau = \tau_b := \inf \{ n \geq 0:  S_n - H({\bm W}_n+n\epsilon_n) > b\},
\end{equation}
where $\epsilon_n \in {\bf R}^d$ with $\epsilon_n \rightarrow 0_d$ as $n \to
\infty$, and $H: {\bf R}^d \rightarrow {\bf R}$ with $H(\vec{0}) = 0$. The
$\epsilon_n$ captures potential additional perturbations.

To simplify the presentation, we refer the reader to \cite{FuhKao-SIAMFM21} for the complete version of these renewal theory results, and restrict ourselves here to the necessary special cases. The first result states that, under suitable regularity conditions and normalization, $({\bm W}_\tau,\tau)$ is asymptotically distributed as a $(d+1)$-dimensional normal random vector. Intuitively, this asymptotic normality arises from a renewal-type central limit theorem, where the cumulative effect of many small increments leads to a Gaussian limit. To describe this result more precisely, we introduce the following notation for characterizing the covariance matrix of $({\bm W}_\tau,\tau)$. For any $\nu \in \mathbf{R}^d$, define
\begin{align}\label{notation-thm-prob}
\notag
M(\nu) = \left( 
\begin{matrix}
0 & I_d \\
-\mu_X^{-1} & \mu_X^{-1}\nu^t
\end{matrix} 
\right), 
& ~~~
\tilde{\Sigma}^*(\nu) = M(\nu) \Sigma M(\nu)^t.
\end{align}
Here $~^t$ denotes transpose. 
Further let $\mathcal{N}_{b,\nu}$ follow a $(d+1)$-dimensional normal
distribution with mean $(0, \cdots, 0, \Eb[\tau_b])^t$ and covariance matrix
$\tilde{\Sigma}^*(\nu)\Eb[\tau_b]$.

\begin{theorem}\label{thm_prob}
Suppose $(X_1, Y_1)$ follows a $(d+1)$-dimensional normal distribution, and $n
\epsilon_n \rightarrow 0$. 
\begin{enumerate}
    \item[1)] If $H \equiv 0$, then
    \begin{equation*}
        \Bigg\vert \Pb\{ {\bm W}_{\tau_b} \in A, \tau_b \leq m \} - \Pb\left\{ \mathcal{N}_{b,\vec{0}} \in A \otimes (-\infty, m] \right\} \Bigg\vert \xrightarrow{b \rightarrow \infty} 0,
    \end{equation*}
	 uniformly for all $A \subset {\bf R}^{d}$ and positive integer $m$, in
	 which  $A \otimes B := \{(x,y): x \in A, y \in B\}.$
    \item[2)] If $H$ is defined as $H(t^1,\ldots,t^d)= \min_{1 \leq i\leq d} t^i$, then
    \begin{equation*}
        \Bigg\vert \Pb\{{\bm W}_{\tau_b} \in A, \tau_b \leq m \} - \sum_{i=1}^d \Pb\left\{ \mathcal{N}_{b,\vec{e}_i} \in (A \cap A_i) \otimes (-\infty, m] \right\} \Bigg\vert \xrightarrow{b \rightarrow \infty} 0,
    \end{equation*}
	 uniformly for all $A \subset {\bf R}^{d}$ and positive integer $m$, where
	 $\vec{e}_i = (0, \cdots, 0, 1, 0, \cdots 0)$ with $1$ at the $i$-th
	 coordinate position, and
    \begin{equation*}
        A_i = \left\{ y = (y_1, y_2, \cdots, y_d): y_i = \min_{1 \leq j \leq d} y_j \right\}.
    \end{equation*}
\end{enumerate}
\end{theorem}

The characterization for $\mathcal{N}_{b,\nu}$ in Theorem~\ref{thm_prob}
requires the characterization of $\Eb[\tau_b]$, which is covered by the following theorem.

\begin{theorem}\label{thm_exp}
Suppose $(X_1, Y_1)$ follows a $(d+1)$-dimensional normal distribution, and $n
\epsilon_n \rightarrow 0$. 
\begin{enumerate}
    \item[1)] If $H \equiv 0$, then there exists constant $\rho_+ > 0$ such that
    \begin{equation*}\label{expectation-0}
    \mu_X \Eb[\tau_b] - b \xrightarrow{b \rightarrow \infty} \rho_+.
    \end{equation*}
	 \item[2)] If $H$ is defined as $H(t^1,\ldots,t^d)= \min_{1 \leq i\leq d} t^i$, then there exist constants
	 $c_0 \in {\bf R}$ and $\rho_+ > 0$ such that 
    \begin{equation*}\label{expectation-H}
        \mu_X \Eb[\tau_b] - b - c_0 \sqrt{b} \xrightarrow{b \rightarrow \infty} \rho_+.
    \end{equation*}
\end{enumerate}
\end{theorem}

The results remain valid when $(X_1, Y_1)$ satisfies conditions {\rm(C1)} and {\rm(C2)}, 
and when $\epsilon_n$ satisfies condition {\rm(C5)} in the Appendix of \citet{FuhKao-SIAMFM21}; 
see Theorem~A.1 therein for details. These results also extend to more general $H$-functions 
and yield the asymptotic distribution of the overshoot
$
S_\tau - H({\bm W}_\tau + \tau \epsilon_\tau) - b.
$
In particular, once the overshoot distribution is available, the constants $\rho_+$ and $c_0$ 
in Theorem~\ref{thm_exp} become explicitly computable 
(see the Appendix of \citet{FuhKao-SIAMFM21}).

\begin{remark}
There are two challenges for the extra term $H$ in \eqref{tau_general}. First,
$H$ can be a non-linear function such as the $\min$ function in 2) of Theorems \ref{thm_prob} and \ref{thm_exp}. Second, ${\bm W}_n$ is a $d$-dimensional random walk. That is,
this additional term transforms the stopping time problem in \eqref{tau_general}
into a multi-dimensional nonlinear boundary crossing problem, which requires a
{\it multivariate nonlinear} renewal theory. 
\end{remark}

Last, we summarize a nonlinear Markov renewal theory from \cite{Fuh-AS04}. 
For convenience of notation, in the remaining part of this and the next subsection, we let $\{(X_n, S_n),~ n \geq 0\}$
denote a Markov random walk on ${\cal X} \times \mathbf{R}$. That is, let $\{X_n, n\geq 0\}$ be a Markov chain on a general state space
${\cal X}$ with $\sigma$-algebra $\cal A$, which is irreducible
with respect to a maximal irreducibility measure on $({\cal X},\cal A)$
and is aperiodic. Let $S_n = \sum_{k=1}^n \xi_k$ be the additive component, taking values on the real line ${\bm R}$, such that
$\{(X_n,S_n), n\geq 0\}$ is a Markov chain on ${\cal X} \times {\bm R}$
with transition probability
\begin{eqnarray}\label{3.1}
& & \Pb\{(X_{n+1},S_{n+1}) \in A \times (B+s) | (X_n,S_n) = (x,s)\} \\
&=& \Pb\{(X_1,S_1) \in A \times B | (X_0,S_0) = (x,0)\}
= \Pb(x,A \times B),  \nonumber
\end{eqnarray}
for all $x \in {\cal X},~ A \in {\cal A}$ and $B \in
{\cal B}({\bm R})$ (:= Borel $\sigma$-algebra on ${\bm R}$).
The chain $\{(X_n,S_n), n \geq 0 \}$ is called a {\it Markov random walk}.
In this subsection, let $\Pb_\nu~(\Eb_{\nu})$ denote the probability (expectation) under
the initial distribution on $X_0$ being $\nu$.
If $\nu$ is degenerate at $x$, we shall simply write $\Pb_x~(\Eb_x)$ instead of
$\Pb_\nu ~(\Eb_{\nu})$. We assume throughout this subsection that
there exists a stationary probability distribution $\pi$, $\pi(A) = \int \Pb(x,A) \drm\pi(x)$
for all $A \in {\cal A}$ and $\Eb_{\pi} [\xi_1]  >0$.

Let $\{Z_n = S_n + \eta_n, n\ge 0\}$ be a
perturbed Markov random walk in the following sense: $S_n$ is a Markov random walk, $\eta_n$ is
${\cal F}_n$-measurable, where ${\cal F}_n$ is the
$\sigma$-algebra generated by $\{(X_k,S_k), 0\le k \le n\}$, and $\eta_n$ is {\it slowly changing}.  Let
$\{A =A(t;\lambda), \lambda \in \Lambda\}$ be a family of boundary
functions for some index set $\Lambda$. Define
\begin{eqnarray}
 T = T_{\lambda} = \inf\{n \geq 1: Z_n > A(n;\lambda) \},~~~\inf \varnothing
 = \infty,~\mbox{for~each}~\lambda \in \Lambda.
\end{eqnarray}
It is easy to see that for all $\lambda > 0$, $T_\lambda < \infty$ with probability $1$.
This section concerns the approximations of the distribution of the overshoot
and expected stopping time $\Eb_\nu [T]$ as the boundary tends to infinity.

A Markov chain $\{X_n,n \geq 0\}$ on a state space ${\cal X}$ is called $V$-uniformly ergodic if
there exists a measurable function $V: {\cal X} \rightarrow [1,\infty)$, with $\int V(x)\drm\pi(x) < \infty$,
such that, for any Borel measurable function $h$ on ${\cal X}$ satisfying
$||h||_V:= \sup_x |h(x)|/V(x)  < \infty$, we have
\begin{eqnarray}
&~& \lim_{n \rightarrow \infty} \sup_{x \in {\cal X}}
\bigg\{ \frac{|\Eb[h(X_n)|X_0=x] - \int h(x)\drm\pi(x)|}{V(x)}
: x \in {\cal X}, |h| \leq V \bigg\} = 0. \label{ue} 
\end{eqnarray}
In this subsection, we shall assume that $\{X_n,n \geq 0\}$ is $V$-uniformly
ergodic. Under irreducibility and aperiodicity assumption, $V$-uniform ergodicity
implies that there exist $ r> 0$ and $0 < \rho <1$ such that for all
$h$ and $n \geq 1$,
\begin{eqnarray}
 \sup_{x \in {\cal X}} \frac{|\Eb[h(X_n)|X_0=x ]- \int h(y) \drm \pi(y)|}
{V(x)} \leq r {\rho}^n \|h\|_V;
\end{eqnarray}
see pages 382-383 of \cite{MeynTweedie2009}.
When $V \equiv 1$, this reduces to the classical uniform ergodicity condition.

The following assumptions for Markov chains will be used in this subsection.

A1. $\sup_x \big\{\frac{\Eb[V(X_1)]}{V(x)} \big\} < \infty$,

A2. $\sup_x \Eb_x [|\xi_1|^2] < \infty$ and
$\sup_x \big\{\frac{\Eb[|\xi_1|^r V(X_1)]}{V(x)} \big\} < \infty$ for some $r \geq 1$.

A3. Let $\nu$ be an initial distribution of the Markov
chain $\{X_n, n \geq 0\}$, assume that for some $r \geq 1$,
\begin{equation}
 \sup_{||h||_V \leq 1} |\int_{x \in {\cal X}} h(x) \Eb_x[|\xi_1|^{r}] \drm\nu(x) |  < \infty.
\end{equation}

A Markov random walk is called {\it lattice} with span $d > 0$ if $d$ is
the maximal number for which there exists a measurable function $\gamma:
{\cal X} \to [0,\infty)$ called the shift function, such that
$\Pb\{\xi_1-\gamma(x) + \gamma(y) \in \{\cdots, -2d, -d, 0, d,
 2d,\cdots\}| X_0 = x, X_1 = y\} = 1$
for almost all $x,y\in {\cal X}$. If no such $d$ exists,
the Markov random walk is called {\it nonlattice}.  A lattice random walk
whose shift function $\gamma$ is identically 0 is called {\it arithmetic}.

To establish the nonlinear Markov renewal theorem, we shall make use of
(\ref{3.1}) in conjunction with the following extension
of Cramer's (strongly nonlattice) condition:
there exists $\delta > 0$ such that for all $m,n=1,2,\cdots$, $\delta^{-1} <m< n$,
and all $\theta \in R$ with $|\theta| \geq \delta$
\begin{eqnarray*}
\Eb_{\pi} | \Eb[\exp(i\theta  (\xi_{n-m}+\cdots+\xi_{n+m}))|X_{n-m},\cdots,X_{n-1},X_{n+1},\cdots,X_{n+m},X_{n+m+1}]|
 \leq e^{-\delta}.
\end{eqnarray*}

By using Markov renewal theory \cite{FuhLai-AAP01,Fuh-AAP04} 
together with Wald's equations for Markov random walks \cite{FuhLai-JAP98}, 
our approach investigates the difference between $T_{\lambda}$ and a stopping 
time defined by crossing linear boundaries with varying drift. Specifically, we define
\begin{equation}\label{tau-def}
   \tau(c,u) = \inf\{n \geq 1 : S_n - un > c\}, 
   \quad c \geq 0,~ u \leq \Eb_\pi [\xi_1],
\end{equation}
and establish the uniform integrability of $|T_{\lambda} - \tau(c_{\lambda},d_{\lambda})|^p$ for $p \geq 1$
for suitable choices of $c_{\lambda}$ and $d_{\lambda}$. 
Nonlinear Markov renewal theory is then derived directly from the 
corresponding results in the linear case, by leveraging uniform integrability 
and the weak convergence of the overshoot.

Let $\Pb_+^u(x,B \times R) = \Pb_x\{X_{\tau(0,u)} \in B \}$ for
$u \leq \Eb_\pi [\xi_1]$, denote the transition probability associated with the Markov
random walk generated by the ascending ladder variable
$S_{\tau(0,u)}$.  Under the $V$-uniform ergodicity condition
and $\Eb_\pi [\xi_1]> 0$, a similar argument as on page 255 of \cite{FuhLai-AAP01} yields that the transition probability $\Pb_+^u(x, \cdot \times R)$ has an invariant measure
$\pi_+^u$. Let $\Eb_{+}^u$ denote expectation under $X_0$ having the initial
distribution $\pi_+^u$. When $u=\Eb_\pi [\xi_1]$, we denote $\Pb_+^{E_\pi \xi_1}$ as $\Pb_+$, and
$\tau_+ = \tau(0,\Eb_\pi [\xi_1])$. Define
\begin{eqnarray}
b &=& b_\lambda = \sup\{t\ge 1: A(t,\lambda) \ge  t \Eb_\pi [\xi_1]\},~~~
 \sup\varnothing = 1, \\
d &=& d_\lambda = (\frac{\partial A}{\partial t} )(b_\lambda;\lambda), \\
\overline{d} &=& \sup\{(\frac{\partial A}{\partial t} )(t;\lambda);~
t\ge b_\lambda,~ \lambda\in\Lambda\},  \\
R &=& R_\lambda = Z_T - A(T; \lambda), \\
R(c,u) &=& S_{\tau(c,u)} - u\tau(c,u) - c,~~~u\le \Eb_\pi [\xi_1],~ c\ge 0, \\
r(u) &=& \Eb_{+}^u [R^2(0,u)]/2\Eb_{_+}^{u} [R(0,u)],~~~ u\le \Eb_\pi [\xi_1], \\
G(r,u) &=& \int_r^\infty \Pb_{+}^u \{R(0,u) > s\}\drm s/
   \Eb_{+}^u [R(0,u)],~~~ u\le \Eb_\pi [\xi_1],~ r\ge 0.
\end{eqnarray}

We shall assume that $A(t;\lambda)$ is twice differentiable in $t$ and
$b_\lambda$ is finite so that $d$ and $\overline{d}$ are well defined.
The next theorem is a Blackwell-type nonlinear Markov renewal theorem.

\begin{theorem}
Assume A1 holds, and A2, A3 hold with $r=1$.
Let $\nu$ be an initial distribution on $X_0$.
Suppose there exist functions $\rho(\delta) > 0$, 
$\sqrt{b} \le \gamma(b) \le b$, $\gamma(b)/b \to 0$ as $b \to \infty$,
and a constant $d^* < \Eb_\pi[\xi_1] \in (0,\infty)$ such that
\begin{align}
\frac{T_\lambda - b_\lambda}{\gamma(b_\lambda)} &= O_{P_\nu}(1),
   && \text{as } b_\lambda \to \infty, \label{eq:th1}\\[4pt]
\lim_{n \to \infty}\!
   \Pb_\nu\!\Big\{ \max_{1 \le j \le \rho(\delta)\gamma(n)}
      |\eta_{n+j} - \eta_n| \ge \delta \Big\} &= 0,
   && \text{for any } \delta > 0, \label{eq:th2}\\[4pt]
\sup\Big\{ \big|\gamma^2(b)\tfrac{\partial^2 A}{\partial t^2}(t;\lambda)\big| :
   |t-b| \le K\gamma(b),~ \lambda\in\Lambda \Big\} &< \infty,
   && \text{for all } K > 0. \label{eq:th3}
\end{align}
Moreover,
\begin{equation}\label{eq:th4}
\lim_{b \to \infty} d_\lambda = d^*.
\end{equation}

If $\xi_1 - d^*$ does not have an arithmetic distribution under $\Pb_\nu$, then for any $r \ge 0$,
\begin{align}\label{eq:th5}
\Pb_\nu\{X_T\in B,\, R_\lambda > r\}
&= \frac{1}{\Eb_{+}^{d^*}[R(0,d^*)]}
    \int_{x\in B} \drm\pi_{+}^{d^*}(x)\,
    \int_r^\infty \Pb_{+}^{d^*}\{R(0,d^*) > s\}\,\drm s \nonumber\\
&\quad + o(1), \qquad \text{as } b_\lambda \to \infty.
\end{align}
In particular, $\Pb_\nu\{R_\lambda > r\} = G(r,d^*) + o(1)$, as  $b_\lambda \to \infty$,
for any $r \ge 0$.

If, in addition, $(T_\lambda - b_\lambda)/\gamma(b_\lambda)$ converges in distribution
to a random variable $W$ as $b_\lambda \to \infty$, then
\begin{equation}\label{eq:th7}
\lim_{b_\lambda \to \infty} 
   \Pb_\nu\{R_\lambda > r,\, T_\lambda \ge b_\lambda + t\gamma(b_\lambda)\}
   = G(r,d^*) \, \Pb_{+}^{d^*}\{W \ge t\},
\end{equation}
for every real $t$ with $\Pb_+^{d^*}\{W=t\} = 0$.
\end{theorem}

To study uniform integrabilitiy of the powers of the differences for linear and
nonlinear stopping times, we shall first give the regularity conditions on $\eta= \{\eta_n, n\ge 1\}$.
The process $\eta $ is said to be {\it regular} with
$p\ge 0$ and $1/2 < \alpha \le 1$ if there exists a random variable
$L$, a function $f(\cdot)$ and a sequence of random variables $U_n,~ n\ge 1$, such that
\begin{eqnarray}
&~& \eta_n = f(n)+U_n, ~\mbox{ for~} n\ge L~ \mbox{ and~} \sup_{x \in {\cal X}} \Eb_x [L^p] <\infty,  \\
&~& \max_{1\le j\le\sqrt n} |f(n+j) - f(n)| \le K,~~~ K < \infty,  \\
&~& \big\{ \max_{1\le j\le n^\alpha} |U_{n+j}|^p,~ n\ge 1 \big\}~
\mbox{is~uniformly~integrable},  \\
&~& n^p \sup_{x \in {\cal X}} \Pb_x \big \{\max_{0\le j\le n} U_{n+j} \ge \theta n^\alpha \big\}
\to 0~~ \mbox{as~} n\to \infty,~{\rm for~all}~\theta > 0,
\end{eqnarray}
and for some $w > 0,~ w < \Eb_\pi [\xi_1] - \overline{d}~ \mbox{if~}\alpha = 1$,
\begin{eqnarray}
 \sum_{n=1}^\infty n^{p-1}\sup_{x \in {\cal X}} \Pb_x\{ -U_n \ge w n^\alpha\} < \infty.
\end{eqnarray}

We shall set $f(n)$ to be the median of $\eta_n$ when $\eta$ is not
regular and extend $f$ to a function on $[1,\infty)$ by linear interpolation.
Therefore we can define $\tau = \tau_\lambda = \tau(c_\lambda, d_\lambda)$
and $ c_\lambda = b_\lambda(\Eb_\pi [\xi_1] - d_\lambda) - f(b_\lambda).$
\begin{theorem}
Assume {\rm A1} holds, and {\rm A2}, {\rm A3} hold with $r=p'(p+1)/\alpha$ for some $ p \geq 1$, $p' >1$
and $1/2 < \alpha \le 1$. Suppose $\eta$ is regular with $p\ge 1$, $1/2 < \alpha \le 1$, and that there exist constants
$\delta$ and $\mu^*$ with $0 < \delta < 1$ and $0< \mu^* < E_\pi [\xi_1]$ such that
\begin{eqnarray}
b^p \sup_x \Pb_x\{T \le \delta b\} \to 0,~~~ {\rm as}~ b\to\infty,
\end{eqnarray}
and
\begin{eqnarray}
\bigg(\frac{\partial A}{\partial t}\bigg)(t;\lambda) \le \mu^*,
~~~ t\ge \delta b,~ \lambda\in\Lambda.
\end{eqnarray}

(i)  If $\sup_{x \in {\cal X}} \Eb_x [ |\xi_1|^{2pp'}] < \infty$ for some $p' >1$
and for any $K > 0$,

\begin{eqnarray}
\sup \{|b_\lambda(\partial^2 A/\partial t^2)(t;\lambda)|:
b_\lambda - Kb_\lambda^\alpha \le t \le b_\lambda + Kb_\lambda^\alpha,~
\lambda\in\Lambda\} < \infty,
\end{eqnarray}
then
\begin{eqnarray}
 \{|T_\lambda - \tau_\lambda|^p;~ \lambda\in\Lambda\}~~~is~
 uniformly~ integrable~under ~\Pb_\nu.
\end{eqnarray}

(ii) If $\partial^2 A/\partial t^2 = 0$, then {\rm (3.27)}
still holds without the condition $\sup_x \Eb_x [|\xi_1|^{2pp'}] < \infty$.
\end{theorem}

We need the following notations and definitions before Theorem~\ref{Thm11}.  

For a given Markov random walk $\{(X_n,S_n), n \geq 0\}$, let $\nu$ be an initial distribution of $X_0$, and define
\[
\nu^*(B) = \sum_{n=0}^\infty \Pb_\nu(X_n \in B), \quad B \in \mathcal{A}.
\]

Let $g = \Eb[\xi_1 \mid X_0, X_1]$ with $\Eb_\pi[|g|] < \infty$.  
Define operators ${\bf P}$ and ${\bf P}_\pi$ by
\[
({\bf P} g)(x) = \Eb_x[g(x,X_1,\xi_1)], \qquad
{\bf P}_\pi g = \Eb_\pi[g(X_0,X_1,\xi_1)],
\]
and set $\overline{g} = {\bf P} g$.  

We shall consider solutions $\Delta(x) = \Delta(x; g)$ of the Poisson equation
\begin{equation}\label{3.28}
\big(I - {\bf P}\big) \Delta = \big(I - {\bf P}_\pi \big) \overline{g}, 
\quad \nu^*\text{-a.s.}, \qquad {\bf P}_\pi \Delta = 0,
\end{equation}
where $I$ is the identity operator.  

Under conditions A1--A4, it is known \cite[Theorem 17.4.2]{MeynTweedie2009} 
that the solution $\Delta$ of \eqref{3.28} exists and is bounded.

\begin{theorem}\label{Thm11}
Assume {\rm A1} holds, and {\rm A2}, {\rm A3} hold with $r=2+p$ for some $p >1$.
Let $\nu$ be an initial distribution such that $\Eb_\nu [V(X_0)] < \infty$. Suppose that
\begin{eqnarray}
&~& \lim_{n\to\infty} \sup_{x\in {\cal X}} \Pb_x \{\max_{1\le j\le\sqrt n} |\eta_{n+j} - \eta_j|
\ge\delta\} = 0 ~~~ {\rm for~ any}~ \delta > 0, \\
&~& \eta_n = f(n) + U_n~~ {\rm for~ any}~ n \ge L,
\end{eqnarray}
and that there exist constants $d_1^* < \Eb_\pi [\xi_1]$ and $d_2^*$ such that
\begin{eqnarray}
&~& \lim_{n\to\infty} \max_{0\le j\le\sqrt n}
     |f(n+j) - f(n)| = 0, \\
&~& U_n~converges~ in~ distribution~ to ~an ~integrable~ random~
variable~U, \\
&~& \lim_{b\to\infty} d_\lambda = d_1^*~ {\rm and}~
  \xi_1 - d_1^*~ is~ nonarithmetic~under ~P_{\nu},
\end{eqnarray}
and for any constant $K > 0$,
\begin{eqnarray}
 \lim_{b\to\infty} \sup\bigg\{\bigg|b_\lambda\bigg(
      \frac{\partial^2 A}{\partial t^2} \bigg)(t;\lambda) - d_2^*\bigg|:
              (t-b_\lambda)^2 \le Kb_\lambda\bigg\} = 0.
\end{eqnarray}
If $\{|T_\lambda - \tau_\lambda|;\lambda\in\Lambda\}$ is uniformly integrable, then
\begin{eqnarray}
\Eb_\nu [T_\lambda]
= b_\lambda - (\Eb_\pi [\xi_1] - d_\lambda)^{-1} f(b_\lambda) + C_0 + o(1),~~~~ as~
b_\lambda\to\infty,
\end{eqnarray}
where
\begin{equation}
\begin{array}{ll}
 C_0
= (\Eb_\pi [\xi_1]-d_1^*)^{-1}\bigg(r(d_1^*)  + (\Eb_\pi [\xi_1] - d_1^*)^{-2} d_2^*\sigma^2/2 - \Eb_{\pi}[U] \\
~~~~~~~~~~~~~~~~~~~~~~~~~~ - \int\Delta(x)\drm(\pi_+^{d^*}(x) - \nu(x))\bigg). \nonumber
\end{array}
\end{equation}
\end{theorem}

When $A(t,\lambda)= \lambda$, we have
\begin{corollary}
Under the assumptions of {\rm Theorem \ref{Thm11}}, as $\lambda \to \infty$
\begin{equation}
\begin{array}{ll}
 \Eb_\nu [T_\lambda] = (\Eb_\pi [\xi_1])^{-1} \bigg(\lambda + \Eb_{\pi_+} [S_{\tau_+}^2]/{2  \Eb_{\pi_+}} [S_{\tau_+}]- f(\lambda/E_\pi \xi_1)  - \Eb_{\pi}[U] \\
 ~~~~~~~~~~~~~~~~~~~~~~~ - \int\Delta(x)d(\pi_+(x)  - \nu(x))\bigg)  + o(1). \nonumber
\end{array}
\end{equation}
\end{corollary}

\subsection{Multivariate Markov Renewal Theory}\label{sec2}

In this subsection, we consider the multivariate Markov renewal theory developed in \cite{FuhLai-AAP01}. For a comprehensive background on multivariate renewal theory, the reader is referred to that paper and the references therein. We begin by presenting multivariate Markov renewal theorems with convergence rates. Recall that $\{X_n, n \geq 0\}$ is assumed to be irreducible (with respect to some measure on ${\cal A}$), aperiodic, and $V$-uniformly ergodic, and
$\{(X_n, S_n := \sum_{k=1}^n \xi_k),~ n \geq 0\}$
is a Markov random walk on ${\cal X} \times \mathbf{R}^d$. 
In addition, we assume 
\begin{eqnarray}
&~& \sup_x\{\frac{\Eb_x[V(X_1)]}{V(x)} \} < \infty, \label{momentcond1} \\ &~& \sup_x \Eb_x[|\xi_1|^2] < \infty~{\rm and~}
\sup_x \Big\{ \frac{\Eb_x[|\xi_1|^r V(X_1)]}{V(x)} \Big\} < \infty. ~~{\rm for~ some}~ r \geq 2. \label{momentcond}
\end{eqnarray}

Let $\pi$ be the stationary distribution of $\{X_n, n
\geq 0\}$ and let $\Pb_\pi$ denote $\int \Pb_x \drm \pi(x)$ and $\Eb_\pi$ be
expectation under $\Pb_\pi$.  Hereafter, we use column
vectors to denote $\theta \in {\bf R}^d$, $\theta^t$ to denote the
transpose of $\theta$, and $|\theta|$ to denote its Euclidean norm
$(\theta^t\theta)^{1/2}$.

Let $\mu = \Eb_\pi[ \xi_1]$ and $\Sigma = \lim_{n \to \infty} n^{-1} \Eb_\pi
[\{(S_n-n\mu)(S_n- n\mu)^t\}]$, which are well defined under (\ref{ue}), 
\eqref{momentcond1} and (\ref{momentcond}). Let $S_{n,j}$ (or $\xi_{n,j}$, $\mu_j$, $\theta_j$) denote the $j$th component of the $d$-dimensional vector $S_{n}$ (or $\xi_{n}$,
$\mu$, $\theta$). Suppose $\mu_1 > 0$. Without loss of generality, it
will be assumed that $\Sigma$ is positive definite (i.e., $\xi_n$ is
strictly $d$-dimensional under $\pi$), because otherwise we can
consider a lower-dimensional subspace instead. In the case $d > 1$ define
\begin{equation}
\gamma =\Eb_\pi[\{(\xi_{n,2}/\mu_1,\cdots,\xi_{n,d}/\mu_1)^t\}],~~~
\tilde{\Sigma} = (-\gamma, \ I_{d-1})\Sigma \left(\begin{array}{c} -\gamma^t
    \\I_{d-1}
\end{array} \right),
\end{equation}
where $I_k$ is the $k\times k$ identity matrix. Note that $\tilde{V}$
is the asymptotic covariance matrix (under $\Pb_\pi$) of
$\{(S_{n,2},\cdots,S_{n,d})^t - S_{n,1} \gamma\} / \sqrt n$. For $s \in
{\bf R}^d$, define $\tilde{s}=(s_2,\cdots,s_d)^t - s_1\gamma$.

First consider the case of i.i.d. $\xi_n$, with $d > 1$ and $S_0=0$.
The renewal measure is defined by $U(B)=\sum_{n=0}^{\infty} \Pb\{S_n\in
B\},$ and multivariate renewal theory is concerned with approximating
$U(s + \cdot)$ by $\Psi_k(s + \cdot)$ as $s_1 \to \infty$, where
$\Psi_k$ is a $\sigma$-finite measure on ${\bf R}^d$ whose density
function (i.e., Radon-Nikodym derivative) with respect to Lebesgue
measure is of the form
\begin{eqnarray}\label{psi}
 \psi_k(s) =
 \frac{1}{\mu_1 \sqrt{\mbox{det} \tilde{\Sigma}}} (\frac{\mu_1}{2 \pi
 s_1})^{(d-1)/2} e^{-\mu_1 \tilde{s}^t \tilde{\Sigma}^{-1} \tilde{s}/2s_1}
 \{1 + \sum_{j=1}^k s_1^{-j/2} \omega_j(\tilde{s}/\sqrt{s_1})\}
\end{eqnarray}
for $s_1 > 0$, and $\psi_k(s)=0$ for $s_1 \leq 0$, where $\omega_j(u)
= \sum_{l=0}^{n_j} q_l(u)$ and $q_l(u)$ is a polynomial of degree $l$
in $u$ whose coefficients are associated with the Taylor expansion of
$(1- \Eb[ e^{i \theta^t \xi_1}])^{-1}$ near $\theta = 0$. 
For Markov random walks, the renewal measure involves not only $\{S_n\}$ but also
$\{X_n\}$. For $A \in {\cal A}$ and $B \in {\cal B}$, define
\begin{eqnarray}\label{RM}
U_\nu^A(B)= \sum_{n=0}^{\infty} \Pb_\nu\{X_n\in A,S_n\in B \}.
\end{eqnarray}
We can approximate $U_\nu^A(s + \cdot)$ by $\pi(A) \Psi_k^{A,\nu}(s +
\cdot)$, in which $\Psi_k^{A,\nu}$ is a $\sigma$-finite measure on
${\bf R}^d$ with density function $\psi_k^{A,\nu}$ with respect to
Lebesgue measure, where $\psi_k^{A,\nu}(s)=0$ for $s_1 \leq 0$ and
$\psi_k^{A,\nu}(s)$ is given by (\ref{psi}) for $s_1 > 0$, with the
coefficients of the polynomials $\omega_1(\tilde{s}),\cdots,
\omega_k(\tilde{s})$ depending also on $A$ and $\nu$ via Taylor's
expansion of the Fourier transform of $U_\nu^A$ near the origin,
assuming that
\begin{eqnarray}\label{mc}
\Eb_\nu[V(X_1)(1+|S_1|^r)] < \infty
\end{eqnarray}
for some sufficiently large $r$ (depending on $k$). Note that when
$\nu$ is degenerate at $(x,0)$, (\ref{mc}) follows from (\ref{momentcond}). The precise
definition of $\omega_j$ is given in Section 4.1 in \cite{FuhLai-AAP01}, where we also prove
the following multidimensional Markov renewal theorem with bounds on
the remainders in approximating $U_\nu^A(s + \cdot)$ by $\pi(A)
\Psi_k^{A,\nu}(s + \cdot)$ as $s_1 \to \infty$, recalling the
assumption $\mu_1 > 0$.

\begin{theorem} 
Let $k \geq 1$ and let $\{(X_n,S_n), n \geq 0\}$ be a strongly nonlattice Markov random walk 
satisfying {\rm (\ref{ue}), \eqref{momentcond1}, (\ref{momentcond})} and {\rm (\ref{mc})} for some $r$.

\begin{enumerate}
\item[(i)] If $r > k + 5 + \max\{1,(d-1)/2\}$, let $A \in \mathcal{A}$ and $B$ be a 
$d$-dimensional rectangle $B = \prod_{j=1}^d [\alpha_j, \beta_j]$. 
Then, as $s_1 \to \infty$,
\[
U_\nu^A \Biggl(s + 
\begin{pmatrix} 0 \\ s_1 \gamma \end{pmatrix} + B \Biggr)
= \pi(A) \, \Psi_k^{A,\nu}\Biggl( s + 
\begin{pmatrix} 0 \\ s_1 \gamma \end{pmatrix} + B \Biggr)
+ o(s_1^{-(d-1+k)/2}),
\] 
uniformly in $\tilde{s}$.

\item[(ii)] If $r > 3$, let $h > 0$ and $\alpha > 0$, and let $\mathcal{B}_{\alpha}$ 
be the class of all Borel subsets of $\mathbb{R}^{d-1}$ such that 
\[
\int_{(\partial B)^\varepsilon} \exp(-|y|^2/2) \, dy = O(\varepsilon^\alpha) 
\quad \text{as } \varepsilon \downarrow 0,
\] 
where $\partial B$ denotes the boundary of $B$ and $(\partial B)^\varepsilon$ 
its $\varepsilon$-neighborhood. Then, as $s_1 \to \infty$,
\begin{equation*}
U_\nu^A \bigl([s_1,s_1+h] \times \sqrt{s_1} (s_1 \gamma + C)\bigr)
= \pi(A) \, \Psi_1^{A,\nu}\bigl([s_1,s_1+h] \times \sqrt{s_1} (s_1 \gamma + C)\bigr) 
+ o(s_1^{-(1+\delta)/2}),
\end{equation*}
for every $\delta < \min(1,r-3)$, uniformly in $A \in \mathcal{A}$ and $C \in \mathcal{B}_{\alpha}$.
\end{enumerate}
\end{theorem}

For $\varepsilon > 0$ and $f:{\bf R}^d \to {\bf R}$,
define the {\it oscillation function} $\Omega_f(s;\varepsilon) =
\sup\{ |f(s)-f(t)|: |s-t| \leq \varepsilon\}.$ Let ${\cal F}_b$ be the
set of all Borel functions $f:{\bf R}^d \rightarrow [0,1]$ such that
$f(s) = 0$ whenever $s_1 \not\in [b,b+h],$ with fixed $h >0$. 

\begin{theorem} Let $\{(X_n,S_n), n \geq 0\}$ be a strongly
  nonlattice Markov random walk satisfying {\rm
    (\ref{ue}), \eqref{momentcond1}, (\ref{momentcond})}  and {\rm (\ref{mc})} for some $r$.
  
(i) If $r> 3$, let $0 < \delta < \min(1,r-3)$. Then for
  every $\eta > 0$, as $b \to \infty$,
\begin{eqnarray*}
\int f(s) dU_\nu^A(s) = \pi(A) \int f(s) \drm\Psi_1^{A,\nu}(s) +O\Bigg(\int
\Omega_f(s;b^{-\eta}) \drm\Psi_1^{A,\nu}(s)\Bigg) + o(b^{-(1+\delta)/2}) 
\end{eqnarray*}
uniformly in $f \in {\cal F}_b$ and $A \in {\cal A}$.

(ii) Suppose $d=1$ and $r \geq 2$. Then as $b \rightarrow \infty$,
\begin{eqnarray*}
 U_\nu^A([b,b+h]) = \pi(A) h/\mu  + o(b^{-(r-1)} ) 
\end{eqnarray*} 
uniformly in $A \in {\cal A}$.
\end{theorem}

It is known that renewal theorems are often
applied to the ladder random walk. The techniques used by \cite{MeynTweedie2009} to prove the $V$-uniform ergodicity of a rich class of time series and queuing models can also
be applied to show that their ladder random walks indeed satisfy
conditions (\ref{ue}), \eqref{momentcond1}, (\ref{momentcond})  and (\ref{mc}). Recall that the (positive) ladder epoch of a Markov random
walk $\{(X_n,S_n), n \geq 0\}$ taking values in ${\cal X} \times {\bf R}$ is defined by
\begin{eqnarray}\label{ladder}
\tau_+ = \inf\{n \geq 1:S_n > 0\}.
\end{eqnarray}
For $A \in {\cal A}$ and Borel subset $B$ of $(0,\infty)$, define
\begin{eqnarray}\label{lmc}
\Pb_+(x, A \times B) = \Pb\{X_{\tau_+} \in A, S_{\tau_+} \in B|X_0=x \}.
\end{eqnarray}
The kernel $\Pb_+$ is the transition probability kernel of a Markov
random walk that has the ladder chain as the underlying Markov chain
on ${\cal X}$. 

Next, we consider limit theorems for first passage times of Markov random walks
by making use of Markov renewal theory for the ladder
random walk, with kernel (\ref{lmc}) in which $\tau_+$ is defined by (\ref{ladder})
in the case $d=1$ and by $\tau_+ = \inf\{n \geq 1: S_{n,1} > 0\}$ in
the case $d > 1$.  It is assumed throughout this section that
  $\Pb_x(\tau_+ < \infty) =1$ for all $x \in {\cal X}$ and that the
  ladder random walk is strongly nonlattice and satisfies conditions
    (\ref{ue}), \eqref{momentcond1}, (\ref{momentcond}) and \eqref{mc}. Let $\pi_+$ denote the invariant measure of the kernel
$\Pb_+(x, A \times {\bf R}^d)$ which is {\it assumed to be irreducible
  and aperiodic}. Let $\tau_1 = \tau_+, \ \tau_{j+1} = \inf\{n >
\tau_j : S_{n,1} > S_{\tau_{j,1}}\}$ 
and
\begin{eqnarray}
T_b &=& \inf\{n \geq 1:S_{n,1} > b \}, \label{TB} \\
\mu^* &=& \Eb_{\pi_+} [S_{\tau_+}],~\Sigma_+ = \lim_{n \rightarrow \infty}
n^{-1} \Eb_{\pi_+} [(S_{\tau_n}- n \mu^*) (S_{\tau_n}- n \mu^*)^t].
\end{eqnarray}
Define $\gamma_+$ and $\widehat{\Sigma}_+$ as in (2.6) but with
$\pi_+,S_{\tau_n} - S_{\tau_{n-1}}, \Sigma_+$ in place of $\pi,\xi_n$ and
$\Sigma$.  

We begin by analyzing the asymptotic distribution of $(X_{T_b}, S_{T_b})$ for a Markov random walk.
 Let $\sigma_+^2 = \lim_{n \to \infty} n^{-1} \Eb_\pi
[S_{\tau_n,1} - n\mu_1^*]^2$.

\begin{theorem}\label{aex}
Assume that $r=2$ in {\rm (\ref{momentcond})} for the
  ladder random walk.  Then as $b \rightarrow \infty$,
\begin{eqnarray*}
\bigg(X_{T_b}, \ S_{T_b,1} - b,~ \sqrt{\frac{\mu_1^*}{b}} \{(S_{T_b,2},
\cdots, S_{T_b,d}) -b \gamma^t_+ \} \bigg)
\end{eqnarray*}
converges weakly under $\Pb_x$ {\rm (}for every $x \in {\cal X}${\rm )}
to $(X, Y, W)$, where $(X,Y)$ and $W$ are independent, $Y$ is a
positive random variable and $X$ takes values in ${\cal X}$ such that
\begin{eqnarray}
\Pb\{X \in A, Y > y\} = \int_{y}^{\infty} \Pb_{\pi_+}\{X_{\tau_+} \in A,
S_{\tau_+,1} > u \} \drm u/\mu_1^*
\end{eqnarray}
for every $A \in {\cal A}$ and $y > 0$, and $W$ is a
$(d-1)$-dimensional Gaussian vector with mean $0$ and covariance
matrix $\widehat{\sigma}_+$.
\end{theorem}

For applications to nonlinear first passage problems, we replace $S_n$
in \eqref{TB} and in Theorem \ref{aex} by $R_n = S_n + \Delta_n$, where $\Delta_n$
represents some nonlinear perturbation. For the case of i.i.d.
increments $\xi_n$ and $d=1$, such extension has been developed by \cite{LaiSiegmund1977}. The following theorem extends their result to the
Markov case and $d \geq 1$. 

\begin{theorem} \label{aex1}
With the same assumptions as in {\rm Theorem
    \ref{aex}}, let ${\cal F}_n$ be the $\sigma$-field generated by
  $\{(X_i,Y_i), 0 \leq i \leq n\}$. Let $\Delta_n$ be ${\cal
    F}_n$-measurable. Assume that for every $x \in {\cal X}$,
\begin{eqnarray}
\max_{1 \leq t \leq n} |\Delta_{t,1}|/n
\stackrel{P_x}{\longrightarrow} 0,~~~\max_{1 \leq t \leq n,2 \leq j
  \leq d} |\Delta_{t,j}|/\sqrt{n} \stackrel{P_x}{\longrightarrow} 0,
\end{eqnarray}
and that for every $\eta > 0$, there exist $\delta = \delta(
\eta,x)$ and $m = m(\eta,x)$ such that
\begin{eqnarray}
\Pb_x\{\max_{n \leq t \leq n + \delta n} |\Delta_{t,1}- \Delta_{n,1}|
\geq \eta\} < \eta~~~for~all ~n \geq m.
\end{eqnarray}
Define $\mu^*,\Sigma_+$ by {\rm (3.2)}. Let $R_n = S_n + \Delta_n$ and
define $T_b^\ast = \inf\{n \geq 1: R_{n,1} \geq b\}$. Then the
conclusion of {\rm Theorem \ref{aex}} still holds with
$(T_b^\ast,R_{T_b^\ast})$ in place of $(T_b,S_{T_b})$. 
\end{theorem}

We next describe an important class of examples that motivate Theorem \ref{aex1}. Let $(X_n,S^*_n)$ be a Markov random walk such that $X_n$ has
stationary distribution $\pi$ and $\mu_{\pi}=\Eb_\pi [S_1^*] \in {\bf
  R}^k$. Suppose $g:{\bf R}^k \to {\bf R}$ and $h:{\bf R}^k \to {\bf
  R}^{d-1}$ are twice continuously differentiable in some neighborhood
of $\mu_\pi$. Let $R_{n,1} = n g(S_n^*/n),~(R_{n,2},\cdots,R_{n,d})^t =
n h(S_n^\ast/n).$ Then Taylor expansions of $g$ and $h$ show that
$R_n$ can be expressed as $S_n + \Delta_n$, where $S_{n,1} =
Dg(\mu_\pi)(S_n^* - n\mu_\pi)$ and $(S_{n,2},\cdots,S_{n,d})^t =
Dh(\mu_\pi)(S_n^* - n\mu_\pi)$, in which $Dg=(\partial g/\partial
s_1,\cdots,\partial g/\partial s_k)$ and $Dh = (\partial h_i/\partial
s_j)_{1 \leq i \leq d-1,1 \leq j \leq k}.$ See \cite{LaiSiegmund1977,LaiSiegmund1979} for certain special cases of $g$ when $S^*_n$ has i.i.d. increments.

Building on these approximations, \cite{FuhLai-AAP01} derive explicit asymptotic expansions 
for the distribution of $(X_{T_b}, S_{T_b})$ as the boundary $b \to \infty$. 
The key idea is to treat the Markov random walk as a ``locally linear'' process 
near the boundary crossing, using the ladder process to capture the first-passage events. 
This approach allows the otherwise complex dependency structure to be expressed 
in terms of the invariant measure $\pi_+$ and the polynomial corrections 
$\omega_1(\tilde{s}/\sqrt{s_1}; x, x_0)$.

For the ladder random walk, the measure $U_+$ and its density $u_+(x;x_0,B)$ play a central role. Intuitively, $u_+(x;x_0,B)$ describes the distribution of the state when the random walk first exceeds the boundary in the positive direction. The approximation
$u_+(x;x_0,B) \approx p_+(x) \, \psi_1^{x,x_0}(s)$ 
provides a tractable way to incorporate the effects of both the initial state $x_0$ and the specific “overshoot” $s$ over the boundary.

These expansions also make it possible to compute functionals of $(X_{T_b}, S_{T_b})$, such as moments or tail probabilities, with controlled error terms. In particular, the convergence rates of the polynomial corrections in $\psi_1^{x,x_0}(s)$ give explicit bounds on the accuracy of the approximations, highlighting the role of the Markov dependence structure and the moment conditions on $\xi_n$.

By expressing the first passage distribution in terms of the ladder process and its invariant measure, one can separate the long-term stationary behavior of the Markov chain from the short-term fluctuations of the random walk. This separation simplifies the analysis and allows classical renewal-theoretic techniques, such as expansions in powers of $1/\sqrt{b}$, to be applied in the multivariate and Markov-dependent setting.

In summary, the multivariate Markov renewal framework provides a powerful tool for approximating the distribution of boundary crossing times and associated states, even in complex dependent settings. The combination of absolute continuity conditions, density approximations, and polynomial corrections offers a flexible and explicit method to handle both finite and general state spaces.

\begin{theorem}
    Suppose $r>3$ in {\rm (\ref{momentcond})} for the ladder
  random walk, which is also assumed to be strongly nonlattice. Let $0
  < \delta < \min(1,r-3)$, $u > 0$ and $\alpha > 0$. Then for every
  $x_0 \in {\cal X}$, as $b \rightarrow \infty$,
\begin{eqnarray*}
&~& \Pb_{x_0} \{X_{T_b} \in A,~S_{T_b,1} >b+u,~ \sqrt{\mu_1^*/b}
((S_{T_b,2},\cdots,S_{T_b,d}) - b \gamma^t_+) \in C \} \\
&=& \frac{1}{\mu_1^*} \int_{\cal X} \bigg\{ \int_{y \in A} 
\int_{s \in {\bf R}^d,s_1 > u} \int_{t = 0}^{s_1 -u} \int_{z \in C + 
\sqrt{\mu_1^*/b}(t\gamma^t_+ - (s_2,\cdots,s_d))}
\frac{ e^{-z^t \widehat{\Sigma}_+^{-1} z/2}}{(2 \pi)^{(d-1)/2} \sqrt{\det 
\widehat{\Sigma}_+}}  \\
&~&  \times [1+\omega_1(z;x,x_0)/\sqrt{b}] \drm z \drm t \Pb_+(x, \drm y \times \drm s) \bigg\}
\drm \pi_+(x) + o(b^{-(1+\delta)/2}) 
\end{eqnarray*}
uniformly in $A \in {\cal A}$ and $C \in {\cal B}_\alpha.$ 
\end{theorem}

If one ignores terms of the order $O(1/\sqrt{b})$ in the integral in
Theorem 7, then the integral reduces via integration by parts to
\begin{eqnarray*}
&~& \Pb \{W \in C\} \int_{\cal X}  \int_{y \in A} 
\int_{s \in {\bf R}^d,s_1 > u} (s_1 -u) \Pb_+(x,\drm y \times \drm s)\drm  \pi_+(x) \\
&=& \Pb \{W \in C\} \int_{u}^{\infty} \Pb_{\pi_+}\{X_{\tau_+} \in A,~
S_{\tau_+,1} > v\}\drm v,
\end{eqnarray*}
where $W$ is a $(d-1)$-dimensional Gaussian vector with mean $0$ and
covariance matrix $\widehat{\Sigma}_+$, as in Theorem \ref{aex}.

In addition to the multivariate Markov renewal theory developed in \cite{FuhLai-AAP01}, 
\cite{FuhLai-JAP98} established Wald's equations, as well as results on first-passage times 
and moments of ladder variables in Markov random walks. Uniform Markov renewal theory 
can be found in \cite{Fuh-AAP04,Fuh-AAP07}. Applications to sequential analysis are discussed 
in \cite{TNB_book2014}, while applications to changepoint detection in hidden Markov models 
appear in \cite{Fuh-AS04, FuhTa2019, Fuh-TIT21}.  
Extending these results to multivariate nonlinear Markov renewal theory remains an interesting and open problem.

\section{SELECTED CONTRIBUTIONS TO BIOSTATISTICS}\label{sec:biostat}

A major theme in the latter portion of Lai's career was applying many of the tools described above from  probability, sequential hypothesis testing, and changepoint detection to designing biomedical clinical trials and other biostatistics problems. In this section we review just a small subset of Lai's biostatistics work, on optimal group sequential designs (Section~\ref{sec:GS}) and on adaptive designs for mid-trial sample size re-estimation (Section~\ref{sec:SS.adj}).  A more complete view of Lai's work in these and related areas can be found in his book \textit{Sequential Experimentation in Clinical Trials: Design and Analysis} \citep{Bartroff13}, whose aim was to provide a bridge from statistical theory to biostatistics practice, as evidenced through the popular R software package \texttt{sp23design} \citep{Narasimhan22} implementing the phase~II/III and other designs covered in the book.  Lai's work in this area also had a sizeable institutional impact, and Lai co-led Stanford Medical School's Biostatistics Core and founded the Center for Innovative Study Design (CISD), producing a stream of adaptive/personalized-medicine trial methodology and applications. Some of Lai's major areas of contribution not covered here are response-adaptive randomization \citep{Lai12}, biomarker-guided and adapative-enrichment designs \citep{Lai13}, comparative-effectiveness and point-of-care (POC) trials \citep{Shih15}, and phase~I dose-finding trials~\citep{Bartroff10b,Bartroff11d}.

\subsection{Group Sequential Testing for Clinical Trials}\label{sec:GS}

The optimality theory discussed in Section~\ref{sec:SHT} for fully-sequential testing would require modification for widespread use in clinical trials, where the dominant methodology is group sequential testing in which groups of patients are analyzed.  \cite{2004power} modified the preceding theory for group sequential tests in a one-parameter
exponential family~\eqref{Expfam} of density
functions, for which Hoeffding's lower bound can be
expressed as
\begin{equation}\label{a13}    
\Eb_\theta(T)\geq-\zeta^{-1}\log(\alpha+\beta)-\left(\zeta^{-2}\sigma/2\right)\left\{(\sigma/4)^2
  -\zeta\log(\alpha+\beta)\right\}^{1/2}+\zeta^{-2}\sigma^2/8
\end{equation}
where
$\sigma^2=(\theta_1-\theta_0)^2 b^{''}(\theta)=\Var_\theta\{(\theta_1-\theta_0)X_i\},\
\zeta=\max\{I(\theta,\theta_0),I(\theta,\theta_1)\}$, and
\begin{equation*}
I(\theta,\lambda)=\Eb_\theta\left[\log\{f_\theta(X_i)/f_\lambda(X_i)\}\right]
=(\theta-\lambda) b'(\theta)-\left( b(\theta)- b(\lambda)\right)
\end{equation*}
is the Kullback--Leibler information number. The lower bound \eqref{a13}
does not take into consideration the fact that $T$ can assume only
several possible values in the case of group sequential designs. The
first step of \citet[p.~509]{2004power} is to take this into
consideration by providing an asymptotic lower bound for $T$ in the following theorem. Let $n_0=0$.
\begin{theorem}
  Suppose the possible values of $T$ are $n_1<\dots<n_k$, such that
\begin{equation}
\lim\inf(n_i-n_{i-1})/|\log(\alpha+\beta)|>0
\label{a14}
\end{equation}
as $\alpha+\beta\to0$, where $\alpha$ and $\beta$ are the type I and
type II error probabilities of the test at $\theta_0$ and $\theta_1$, respectively. Let
$m_{\alpha,\beta}(\theta)=\min\{|\log\alpha|/I(\theta,\theta_0),|\log\beta|/I(\theta,\theta_1)\}$.
Let $\epsilon_{\alpha,\beta}$ be positive numbers such that
$\epsilon_{\alpha,\beta}\to0$ as $\alpha+\beta\to0$, and let $\nu$ be
the smallest $j(\leq k)$ such that $n_j\geq(1-\epsilon_{\alpha,\beta})m
_{\alpha,\beta}(\theta)$, defining $\nu$ to be $k$ if no such $j$
exists. Then for fixed $\theta,\theta_0$ and $\theta_1>\theta_0$, as
$\alpha+\beta\to0$,
$\Pb_\theta(T\geq n_\nu)\to1$.
If furthermore $\nu<k,\ |m_{\alpha,\beta}(\theta)-n_\nu|/m_{\alpha,\beta}^{1/2}(\theta)\to0$
and
\begin{equation*}
\lim\sup\frac{m_{\alpha,\beta}(\theta)}{\max\left\{|\log\alpha|/I(\theta,\theta_0),\ |\log\beta|/I(\theta,\theta_1)\right\}}<1,
\label{a15}
\end{equation*}
then $\Pb_\theta(T\geq n_{\nu+1})\geq\frac12+o(1)$.
\label{thma1}
\end{theorem}

The $n_j$ in Theorem~\ref{thma1} can in fact be random variables independent
of $X_1,X_2,\dots$. In this case the preceding
argument can still be applied after conditioning on $(n_1,\dots,n_k)$. The next step of
\citet[p.~510]{2004power} is to extend Theorem~\ref{Th:AO2SPRTnoniid} about asymptotic optimality of Lorden's 2-SPRT to the group sequential setting in
the following. Let $S_n=\sum_{i=1}^n X_i$.
\begin{theorem}\label{thma2}
Let $\theta_0<\theta^*<\theta_1$ be such that
  $I(\theta^*,\theta_0)=I(\theta^*,\theta_1)$. Let $\alpha+\beta\to0$
  such that $\log\alpha\sim\log\beta$.
\begin{enumerate}
\item The sample size $n^*$ of the Neyman--Pearson test of
  $\theta_0$ versus $\theta_1$ with error probabilities $\alpha$ and
  $\beta$ satisfies $n^*\sim|\log\alpha|/I(\theta^*,\theta_0)$.
\item For $L\geq1$, let $\mathcal{T}_{\alpha,\beta,L}$ be the class of
  stopping times associated with group sequential tests with error
  probabilities not exceeding $\alpha$ and $\beta$ at $\theta_0$ and
  $\theta_1$ and with $k$ groups and prespecified group sizes such
  that \eqref{a14} holds and $n_k=n^*+L$. Then, for given $\theta$ and
  $L$, there exists $\tau\in\mathcal{T}_{\alpha,\beta,L}$ that stops
  sampling when
  \begin{equation*}(\theta-\theta_0)S_{n_i}-n_i\left\{ b(\theta)- b(\theta_0)\right\}\geq b\quad\mbox{or}\quad(\theta-\theta_1)S_{n_i}-n_i\left\{ b(\theta)- b(\theta_1)\right\}\geq\tilde{b}
  \end{equation*}
  for $1\leq i\leq k-1$, with $b\sim|\log\alpha|\sim\tilde{b}$, and
  such that
  \begin{equation}
  \Eb_\theta(\tau)\sim\inf_{T\in\mathcal{T}_{\alpha,\beta,L}}\Eb_\theta(T)\sim n_\nu+\rho(\theta)(n_{\nu+1}-n_\nu),
  \label{a17}
  \end{equation}
  where $\nu$ and $m_{\alpha,\beta}(\theta)$ are defined in Theorem~\ref{thma1} and $0\leq\rho(\theta)\leq1$.
\end{enumerate}
\end{theorem}

Whereas the theory for the group sequential 2-SPRT in Theorem~\ref{thma2} requires
specification of $\theta$, the group sequential GLR in the next section
replaces $\theta$ at the $i$th interim analysis by the maximum
likelihood estimate $\hat\theta_{n_i}$, similar to the tests in Section~\ref{ssec:GLR} for the fully sequential case.
The resulting group sequential GLR test of \citet[p.~511--512]{2004power} attains the asymptotic lower bound
\eqref{a17} at every fixed $\theta$ and that its power is comparable
to the upper bound $1-\beta$ at $\theta_1$,
under the assumption that the group sizes satisfy \eqref{a14} with
$n_k \sim |\log\alpha|/I(\theta^*,\theta_0)$, as $\alpha+\beta\to0$
such that $\log\alpha\sim\log\beta$.

\subsection{Efficient Adaptive Designs for Mid-Course Sample Size Adjustment}\label{sec:SS.adj}

In this section, we review some of Lai's and his coathors' work on re-estimating the sample size for a group sequential test midway through a clinical trial.  There was increasing interest in this topic because of the ethical and economic considerations in the design of
clinical trials to test the efficacy of new treatments and lack of information on the new treatments being tested.  One challenge in this area is the difficulty in specifying the alternative on which the power of a test is based, and pilot studies may be unavailable or difficult to interpret, thus there was interest in finding ways to adjust the test statistics while maintaining control on the the type I error probability.

Much of the existing literature focused on 2-stage methods for testing about the normal mean. A unified treatment, developed by \citet{Bartroff08c,Bartroff08} in the general framework
of multiparameter exponential families  uses efficient GLR statistics and adds a third
stage to adjust for the sampling variability of the first-stage
parameter estimates that determine the second-stage sample size. The
possibility of adding a third stage to improve two-stage designs dated
back to \citet{Lorden83}. Whereas Lorden's upper bounds for
the type I error probability are too conservative for clinical trial applications which must follow strict regulatory specifications, \citet{Bartroff08c} overcame this difficulty by
modifying the numerical methods to compute the type I
error probability, and also extended the three-stage test to
multiparameter and multi-armed settings, thus greatly broadening the
scope of these efficient adaptive designs.

\subsubsection{An Adaptive 3-stage GLR Test}\label{sec:adapt}

\citet{Bartroff08c,Bartroff08} consider the  general framework of
the multiparameter exponential family $f_{\btheta}(x)=\exp(\btheta^T x-\psi(\btheta))$ considered in \eqref{dist.exp.family}. Let $\Lambda_{i,j}$, $j\in\{0,1\}$, denote the GLR statistic in this family
comparing $\bm{\widehat{\theta}}_{n_i}$ to the composite~$u(\bm{\theta})=u_j$ and $u$ is a
smooth real-valued function defining the hypotheses, below. Rather than considering  simple null and
alternative hypotheses, \citet{Bartroff08c} use the GLR statistics
$\Lambda_{i,0}$ and $\Lambda_{i,1}$ in an adaptive three-stage test of
the composite null hypothesis $H_0: u(\bm{\theta})\le u_0$,  such that
\begin{equation*}
\text{$I(\bm{\theta},\bm{\lambda})$ is increasing in $|u(\bm{\lambda})-u(\bm{\theta})|$ for every fixed $\bm{\theta}$.}
\end{equation*}
Let $n_1=m$ be the sample size of the first stage and $n_3=M$ be the
maximum total sample size, both specified before the trial. Let
$u_1>u_0$ be the alternative implied by the maximum sample size $M$
and the reference type II error probability $\tilde{\alpha}$. That is,
$u_1(>u_0)$ is the alternative where the fixed sample size (FSS) GLR
test with type I error probability $\alpha$ and sample size $M$ has
power $\inf_{\bm{\theta}:u(\bm{\theta})=u_1} \Pb_{\bm{\theta}}\{\text{Reject $H_0$}\}$
equal to $1-\tilde{\alpha}$. The three-stage test of
$H_0: u(\bm{\theta})\le u_0$ stops and rejects $H_0$ at stage $i\le 2$ if
\begin{equation}\label{eq:6}
n_i<M,\quad u\big(\bm{\widehat{\theta}}_{n_i}\big)>u_0,\quad\text{and}\quad\Lambda_{i,0}\geq b.
\end{equation}
Early stopping for futility (accepting $H_0$) can also occur at stage
$i\leq2$ if
\begin{equation}\label{eq:7}
n_i<M,\quad u\big(\bm{\widehat{\theta}}_{n_i}\big)<u_1,\quad\text{and}\quad\Lambda_{i, 1} \geq \tilde b.
\end{equation}
The test rejects $H_0$ at stage $i=2$ or $3$ if
\begin{equation}\label{eq:8}
n_i=M,\quad u\big(\bm{\widehat{\theta}}_M\big)>u_0,\quad\text{and}\quad\Lambda_{i,0}\geq c,
\end{equation}
accepting $H_0$ otherwise. The sample size $n_2$ of the three-stage
test is given by
\begin{equation*}
n_2=m\vee\left\{M \wedge\left\lceil (1 + \rho_m)\, n\big(\bm{\widehat{\theta}}_m\big)\right\rceil\right\},
\end{equation*}
with
\begin{equation}\label{eq:5}
n(\bm{\theta}) = \min\left\{|\log\alpha|\Big/\inf_{\bm{\lambda}:u(\bm{\lambda})=u_0} I(\bm{\theta},\bm{\lambda}),\
|\log\tilde{\alpha}|\Big/\inf_{\bm{\lambda}:u(\bm{\lambda})=u_1}I(\bm{\theta},\bm{\lambda})\right\},
\end{equation}
where $I(\bm{\theta},\bm{\lambda})$ is the Kullback--Leibler information number
and $\rho_m>0$ is an inflation factor to adjust for
uncertainty in $\bm{\widehat{\theta}}_m$.
Note that \eqref{eq:5} is an asymptotic approximation to Hoeffding's lower bound \eqref{a13}.
Letting $0<\varepsilon,\tilde{\varepsilon}<1$, define the thresholds $b, \tilde{b}$,
and $c$ to satisfy the equations
\begin{align}
&\sup_{\bm{\theta}:u(\bm{\theta})=u_1}\Pb_{\bm{\theta}}\{\text{\eqref{eq:7} occurs for $i=1$ or 2}\}=\tilde\varepsilon\tilde{\alpha},\label{eq:9}\\
&\sup_{\bm{\theta}:u(\bm{\theta})=u_0}\Pb_{\bm{\theta}}\{\text{\eqref{eq:7} does not occur for $i\leq2$, \eqref{eq:6} occurs for $i=1$ or 2}\}=\varepsilon\alpha,\\
&\sup_{\bm{\theta}:u(\bm{\theta})=u_0}\Pb_{\bm{\theta}}\{\text{\eqref{eq:6} and \eqref{eq:7} do not occur for $i\leq2$, \eqref{eq:8} occurs}\}=(1-\varepsilon)\alpha.\label{eq:11}
\end{align}
The probabilities in \eqref{eq:9}--\eqref{eq:11} can be computed by
using the normal approximation to the signed-root likelihood ratio
statistic
\begin{equation*}
\ell_{i,j}=\left\{\text{sign}\left(u(\bm{\widehat{\theta}}_{n_i})-u_j\right)\right\}
(2n_i\Lambda_{i,j})^{1/2}
\end{equation*}
($1\le i\le 3$ and $j=0,1$) under $u(\bm{\theta})=u_j$.
When $u(\bm{\theta})=u_j$, $\ell_{i,j}$ is approximately normal with mean
0, variance $n_i$, and the increments $\ell_{i,j}-\ell_{i-1,j}$ are
asymptotically independent. We can therefore approximate $\ell_{i,j}$
by a sum of independent standard normal random variables under
$u(\theta)=u_j$ and thereby determine $b, \tilde{b}$, and $c$.  Note
that this normal approximation can also be used for the choice of
$u_1$ implied by $M$ and $\tilde{\alpha}$. 

A special multiparameter case of particular interest in clinical
trials involves $K$ independent populations having density functions
$\exp\{\theta_k x-\tilde{\psi}_k(\theta_k)\}$ so that
$\bm{\theta}^T \bm{x}-\psi(\bm{\theta})=\sum_{k=1}^K \{\theta_k x_k-\tilde{\psi}(\theta_k)\}$.
In multi-arm trials, for which different numbers of patients are
assigned to different treatments, the GLR statistic $\Lambda_{i,j}$
for testing the hypothesis $u(\theta_1,\dots,\theta_K)=u_j$ ($j=0$ or
$1$) at stage $i$ has the form
\begin{equation*}
\Lambda_{i,j}=\sum_{k=1}^Kn_{ki}\left\{\widehat{\theta}_{k,n_{ki}}\bar{X}_{k,n_{ki}}-
   \tilde\psi\left(\widehat{\theta}_{k,n_{ki}}\right)\right\}
-\sup_{\bm{\theta}: u(\theta_1,\dots,\theta_K)=u_j}
\,\sum_{k=1}^K n_{ki} \left\{\theta_k\bar{X}_{k,n_{ki}}-\tilde\psi(\theta_k)\right\},
\end{equation*}
in which $n_{ki}$ is the total number of observations from the $k$th
population up to stage $i$. Letting $n_i=\sum_{k=1}^K n_{ki}$, the normal approximation to the
signed root likelihood ratio statistic is still applicable when
$n_{ki}=p_kn_i+O_p(n_i^{1/2})$, where $p_1,\dots,p_K$ are
nonnegative constants that sum up to 1, as in random allocation of
patients to the $K$ treatments (for which $p_k=1/K$); see \citet[p.~514]{2004power}.

\subsubsection{Mid-Course Modification of Maximum Sample Size}\label{sec:Mtild}
The adaptive designs in the preceding section can be modified to
accommodate the possibility of mid-course increase of the maximum
sample size from $M$ to $\widetilde{M}$. Let $u_2$ be the alternative
implied by $\widetilde{M}$ so that the level-$\alpha$ GLR test with sample
size $\widetilde{M}$ has power $1-\tilde{\alpha}$. Note that
$u_1>u_2>u_0$. Whereas the sample size $n_3$ is chosen to be $M$ in
Section~\ref{sec:adapt}, we now define
\begin{align*}
\tilde{n}(\bm{\theta})&=\min\left\{|\log\alpha|\Big/\inf_{\bm{\lambda}:u(\bm{\lambda})=u_0}I(\bm{\theta},\bm{\lambda}),\ |\log\tilde{\alpha}|\Big/\inf_{\bm{\lambda}:u(\bm{\lambda})=u_2}I(\bm{\theta},\bm{\lambda})\right\},\\
n_3&=n_2\vee\left\{M'\wedge\left\lceil(1+\rho_m)\tilde{n}\big(\hat{\bm{\theta}}_{n_2}\big)\right\rceil\right\},
\end{align*}
where $M<M' \le \widetilde{M}$ and $n_2=m\vee \{M\wedge
(1+\rho_m)\tilde{n}(\bm{\widehat{\theta}}_m)\}$. We can regard the test as a
group sequential test with 4 groups and $n_1=m$, $n_4=\widetilde{M}$, but
with adaptively chosen $n_2$ and $n_3$. If the test does not end at
the third stage, continue to the fourth and final stage with sample
size $n_4=\widetilde{M}$. Its rejection and futility boundaries are
similar to those in Section~\ref{sec:adapt}. Extending our notation
$\Lambda_{i,j}$ to $1\le i\le 4$ and $0\le j\le 2$, the test stops at
stage $i\le 3$ and rejects $H_0$ if
\begin{equation*}
n_i<\widetilde{M},\quad
u\big(\bm{\widehat{\theta}}_{n_i}\big)>u_0,\quad\text{and}\quad\Lambda_{i,0}\geq b,
\end{equation*}
stops and accepts $H_0$ if
\begin{equation}\label{eq:15}
n_i<\widetilde{M},\quad u\big(\hat{\bm{\theta}}_{n_i}\big)<u_2,\quad\text{and}\quad\Lambda_{i,2}\geq\tilde{b},
\end{equation}
and rejects $H_0$ at stage $i=3$ or $4$ if
\begin{equation*}
n_i=\widetilde{M},\quad u\big(\hat{\bm{\theta}}_{\widetilde{M}}\big)>u_0,\quad\text{and}\quad
\Lambda_{i,0}\geq c,
\end{equation*}
accepting $H_0$ otherwise. The thresholds $b, \tilde{b}$, and $c$ can
be defined by equations similar to \eqref{eq:9}--\eqref{eq:11} to
insure the overall type I error probability to be $\alpha$. For
example, in place of \eqref{eq:9},
\begin{equation}\label{eq:16}
\sup_{\bm{\theta}:u(\bm{\theta})=u_2} \Pb_\theta \{\text{\eqref{eq:15} occurs for some $i\le
  3$}\}=\tilde{\varepsilon}\tilde{\alpha}.
\end{equation}
The basic idea underlying \eqref{eq:16} is to control the type II
error probability at $u_2$ so that the test does not lose much power
there in comparison with the GLR test that has sample size $\widetilde{M}$
(and therefore power $1-\tilde{\alpha}$ at $u_2$).

\subsubsection{Asymptotic Theory for the Adaptive 3-Stage GLR Test}\label{3sub23new}

\cite{Bartroff08c,Bartroff08} established the asymptotic optimality of the above three-stage test in the following.

\begin{theorem}
Let $N$ denote the sample size of the three-stage GLR test in Section~\ref{sec:adapt}, with $m$, $M$, and $m\vee[M\wedge\lceil(1+\rho_m)n(\hat{\bm{\theta}}_m)\rceil]$ being the possible values of $N$. Let $T$ be the sample size of any test of $H_0:u(\bm{\theta})\leq u_0$ versus $H_1:u(\bm{\theta})\geq u_1$, sequential or otherwise, which takes at least $m$ and at most $M$ observations and whose type I and type II error probabilities do not exceed $\alpha$ and $\tilde{\alpha}$, respectively. Assume that $\log\alpha\sim\log\tilde{\alpha}$,
\begin{equation}
m/|\log\alpha|\to a,\quad M/|\log\alpha|\to A,\quad\rho_m\to0 \ \text{ but }\ m^{1/2}\rho_m/(\log m)^{1/2}\to\infty
\label{4.1}
\end{equation}
as $\alpha+\tilde{\alpha}\to0$, with $0<a<A$. Then for every fixed $\bm{\theta}$, as $\alpha+\tilde{\alpha}\to0$,
\begin{align}
\Eb_{\bm{\theta}}(N)&\sim m \vee \left\{M\wedge|\log\alpha|/ \left[\inf_{\bm{\lambda}:u(\bm{\lambda})=u_0} I(\bm{\theta},\bm{\lambda}) \vee
\inf_{\bm{\lambda}:u(\bm{\lambda})=u_1} I(\bm{\theta},\bm{\lambda})\right]\right\},\label{4.2}\\
\Eb_{\bm{\theta}}(T) & \geq [1+o(1)] \Eb_{\bm{\theta}}(N).\label{4.3}
\end{align}
\label{thm81}
\end{theorem}

Since $m\sim a|\log\alpha|$ and $M\sim A|\log\alpha|$ and since the thresholds $b$, $\tilde{b}$, and $c$
are defined by \eqref{eq:9}--\eqref{eq:11},  \cite{Bartroff08} use an argument similar to
the proof of Theorem 2(ii) of \citet[p.~525]{2004power} to show that \eqref{4.2} holds.
They then use the following argument to prove \eqref{4.3}. Let $\Theta_0=\{\bm{\theta}:u(\bm{\theta})\leq u_0\}$, $\Theta_1=\{\bm{\theta}:u(\bm{\theta})\geq u_1\}$. For $i=0,1$,
\begin{equation}
\inf_{\bm{\lambda} \in \Theta_i}I(\bm{\theta},\bm{\lambda})=I_i(\bm{\theta}),
\qquad\text{where } I_i(\bm{\theta})=\inf_{\bm{\lambda}:u(\bm{\lambda})=u_i} I(\bm{\theta},\bm{\lambda}).
\label{4.4}
\end{equation}
Take any $\bm{\lambda}\in\Theta_0$ and $\tilde{\bm{\lambda}}\in\Theta_1$.
From \eqref{4.4} and Hoeffding's lower bound~\eqref{a13}, it follows that for a test
that has error probabilities $\alpha$ and $\tilde{\alpha}$
at $\bm{\lambda}$ and $\tilde{\bm{\lambda}}$ and take at least $m$ and at most $M$ observations, its sample size $T$ satisfies
\begin{equation}
\Eb_{\bm{\theta}}(T)\geq m\vee\left\{M\wedge\frac{[1+o(1)]|\log\alpha|}{I_0(\bm{\theta})\vee I_1(\bm{\theta})}\right\}
\label{4.5}
\end{equation}
as $\alpha+\tilde{\alpha}\to0$ such that $\log\alpha\sim\log\tilde{\alpha}$. The second-stage sample size of the
adaptive test is a slight inflation of the Hoeffding-type lower bound \eqref{4.5} with $\bm{\theta}$ replaced by
the maximum likelihood estimate $\hat{\bm{\theta}}_m$ at the end of the first stage.
The assumption $\rho_m\to0$ but $\rho_m\succ m^{-1/2}(\log m)^{1/2}$ is used to accommodate
the difference between $\bm{\theta}$ and its substitute $\hat{\bm{\theta}}_m$, which satisfies
$\Pb_{\bm{\theta}}\{\sqrt{m} \| \hat{\bm{\theta}}_m-\bm{\theta} \| \geq r(\log m)^{1/2}\} = o(m^{-1})$
if $m$ is sufficiently large, by standard exponential bounds involving moment generating functions.

As noted by \citet{Bartroff08}, the adaptive test in Section~\ref{sec:Mtild} can be regarded as a mid-course amendment of an adaptive test of $H_0:u(\bm{\theta})\leq u_0$ versus $H_1:u(\bm{\theta})\geq u_1$, with a maximum sample size of $M$, to that of $H_0$ versus $H_2:u(\bm{\theta})\geq u_2$, with a maximum sample size of $\widetilde{M}$. Whereas \eqref{4.5} provides an asymptotic lower bound for tests of $H_0$ versus $H_1$, any test of $H_0$ versus $H_2$ with error probabilities not exceeding $\alpha$ and $\tilde{\alpha}$ and taking at least $m$ and at most $\widetilde{M}$ observations likewise satisfies
\begin{equation}
\Eb_{\bm{\theta}}(T)\geq m\vee\left\{\widetilde{M}\wedge\frac{[1+o(1)]|\log\alpha|}{I_0(\bm{\theta})\vee I_2(\bm{\theta})}\right\}
\label{4.6}
\end{equation}
as $\alpha+\tilde{\alpha}\to0$ such that $\log\alpha\sim\log\tilde{\alpha}$. Note that $\Theta_1=\{\bm{\theta}:u(\bm{\theta})\geq u_1\} \subset\Theta_2=\{\bm{\theta}:u(\bm{\theta})\geq u_2\}$ and therefore $I_2(\bm{\theta})\leq I_1(\bm{\theta})$.
The four-stage test in Section~\ref{sec:Mtild}, with $M'=\widetilde{M}$, attempts to attain the asymptotic lower bound in \eqref{4.5} prior to the third stage and the asymptotic lower bound in \eqref{4.6} afterwards. It replaces $I_1(\bm{\theta})$ in \eqref{4.5}, which corresponds to early stopping for futility, by $I_2(\bm{\theta})$ that corresponds to rejection of $H_2$ (instead of $H_1$) in favor of $H_0$. Thus, the second-stage sample size $n_2$ corresponds to the lower bound in \eqref{4.5} with $\bm{\theta}$ replaced by $\hat{\bm{\theta}}_m$ and $I_1$ replaced by $I_2$, while the third-stage sample size corresponds to that in \eqref{4.6} with $\bm{\theta}$ replaced by $\hat{\bm{\theta}}_{n_2}$. The arguments used to prove the asymptotic optimality of the three-stage test in Theorem~\ref{thm81} are then modified to prove the following.
\begin{theorem}
Let $N^*$ denote the sample size of the four-stage GLR test in Section~\ref{sec:Mtild}, with $M'=\widetilde{M}$. Assume that $\log\alpha\sim\log\tilde{\alpha}$ as $\alpha+\tilde{\alpha}\to0$, that \eqref{4.1} holds and $\widetilde{M}/|\log\alpha|\to\widetilde{A}$ with $0<a<A<\widetilde{A}$. Then
\begin{equation*}
\Eb_{\bm{\theta}}(N^*)\sim\begin{cases}m\vee[1+o(1)]|\log\alpha|/I_0(\bm{\theta})&\text{if }I_0(\bm{\theta})>A^{-1},\\
m\vee\left\{\widetilde{M} \wedge[1+o(1)]|\log\alpha|/[I_0(\bm{\theta})\wedge I_2(\bm{\theta})]\right\}&\text{if }I_0(\bm{\theta})<A^{-1}.\end{cases}
\end{equation*}
\label{thm32}
\end{theorem}

\section*{Acknowledgements}

JB: Lai was my postdoc advisor at Stanford in the ``mid-aughts,'' when I first got to witness his genius and work pace (``The speed of Lai,'' as his Stanford colleague David~Rogosa described it), usually in the Stanford Math Library in the middle of the night. I am grateful for Lai's generosity in our research collaborations, his guidance of my young career, his contagious positivity, and the many laughs we shared along the way. 

CDF: I have greatly enjoyed collaborating with Professor Tze-Liang Lai on Markov renewal theory. I am also deeply grateful to Prof.~Lai for many fruitful conversations between 1990 and 2023, which played a significant role in my research, particularly in the areas of importance sampling and the multi-armed bandit problem.

AT:  I am deeply grateful to Tze Lai for the many insightful conversations we shared between 1993 and 2023, which played a significant role in shaping my work. 
Tze's work meaningfully influenced my research, from 1981 on.

HX: I am truly indebted to my advisor and collaborator, 
Professor Tze Leung Lai, for his guidance and inspiring 
mentorship. Our many discussions and collaboration over 
the years has profoundly shaped my research and the mentoring 
of students.


\begin{thebibliography}{55}
\newcommand{\enquote}[1]{``#1''}
\providecommand{\natexlab}[1]{#1}
\providecommand{\url}[1]{\normalfont{#1}}
\providecommand{\urlprefix}{}


\bibitem[Bartroff and Lai, 2008a]{Bartroff08c}
Bartroff, J. and T. L. Lai. 2008a.
\newblock ``Efficient Adaptive Designs with Mid-Course Sample Size Adjustment in
  Clinical Trials.''
\newblock {\em Statistics in Medicine}, 27:1593--1611.

\bibitem[Bartroff and Lai, 2008b]{Bartroff08}
Bartroff, J. and T. L. Lai . 2008b.
\newblock ``Generalized Likelihood Ratio Statistics and Uncertainty Adjustments
  in Adaptive Design of Clinical Trials.''
\newblock {\em Sequential Analysis}, 27:254--276.

\bibitem[Bartroff and Lai, 2010]{Bartroff10b}
Bartroff, J. and T. L. Lai. 2010.
\newblock ``Approximate Dynamic Programming and its Applications to the Design of
  Phase {I} Cancer Trials.''
\newblock {\em Statistical Science}, 25:245--257.

\bibitem[Bartroff and Lai, 2011]{Bartroff11d}
Bartroff, J. and T. L. Lai. 2011.
\newblock ``Incorporating Individual and Collective Ethics Into Phase {I} Cancer
  Trial Designs.''
\newblock {\em Biometrics}, 67:596--603.

\bibitem[Bartroff, Lai, and Shih(2013)]{Bartroff13}
Bartroff, J., T.~L. Lai, and M.~Shih. 2013. \emph{Sequential Experimentation in Clinical Trials: Design and Analysis}. New York: Springer.

\bibitem[Chen et~al.(2011)Chen, Xing, and Zhang]{ChenXingZhang2011}
H.~Chen, H.~Xing, and N.~Zhang. 2011.
\newblock ``Estimation of Parent Specific DNA Copy Number in Tumors Using High-Density Genotyping Arrays.''
\newblock \emph{PLoS Computational Biology}, 7 (1): e1001060.

  
   \bibitem[Dragalin, Tartakovsky, and Veeravalli(2000)]{DTVPart1_IEEEIT1999}
  Dragalin, V.~P., A.~G. Tartakovsky, and V.~V. Veeravalli. 1999.  
  `` Multihypothesis Sequential Probability Ratio Tests -- Part I: Asymptotic Optimality.''
  \emph{IEEE Transactions on Information Theory} 45(11):2448--2461.
  
  \bibitem[Dragalin, Tartakovsky, and Veeravalli(2000)]{DTVPart2_IEEEIT2000}
  Dragalin, V.~P., A.~G. Tartakovsky, and V.~V. Veeravalli. 2000.  
  `` Multihypothesis Sequential Probability Ratio Tests -- Part II: Accurate Asymptotic Expansions for the  Expected Sample Size.''
  \emph{IEEE Transactions on Information Theory} 46(4):1366--1383.


  \bibitem[Fuh(2004a)]{Fuh-AAP04}
Fuh, C. D. 2004a. ``Uniform {M}arkov Renewal Theory and Ruin Probabilities in {M}arkov Random Walks.'' \emph{The Annals of Applied Probability} 14 (3): 1202--1241.

\bibitem[Fuh(2004b)]{Fuh-AS04}
Fuh, C. D. 2004b. ``Asymptotic Operating Characteristics of an Optimal Change Point Detection in Hidden {M}arkov Models.'' \emph{The Annals of Statistics} 32 (5): 2305--2339.

\bibitem[Fuh(2007)]{Fuh-AAP07}
Fuh, C. D. 2007. ``Asymptotic Expansions on Moments of the First Ladder Height in {M}arkov Random Walks With Small Drift.'' \emph{Advances in Applied Probability} 39 (3): 826--852.

\bibitem[Fuh(2021)]{Fuh-TIT21}
Fuh, C. D. 2021. ``Asymptotically Optimal Change Point Detection for Composite Hypothesis in State Space Models.'' \emph{IEEE Transactions on Information Theory} 67 (1): 485--505.

\bibitem[Fuh and Kao(2021)]{FuhKao-SIAMFM21}
Fuh, C. D. and C. L. Kao. 2021. ``Credit Risk Propagation in Structure Form Models.'' \emph{SIAM Journal on Financial Mathematics} 12 (4): 1340--1373.

\bibitem[Fuh and Lai(1998)]{FuhLai-JAP98}
Fuh, C. D. and T. L. Lai. 1998. ``Wald’s Equations, First Passage Times and Moments of Ladder Variables in {M}arkov Random Walks.'' \emph{Journal of Applied Probability} 35 (3): 566--580.

\bibitem[Fuh and Lai(2001)]{FuhLai-AAP01}
Fuh, C. D. and T. L. Lai. 2001. ``Asymptotic Expansions in Multidimensional {M}arkov Renewal Theory and First Passage Times for {M}arkov Random Walks.'' 
\emph{Advances in Applied Probability} 33 (3): 652--673.


\bibitem[Fuh and Tartakovsky(2019)]{FuhTa2019}
Fuh, C. D. and A. G. Tartakovsky. 2019. ``Asymptotic Bayesian Theory of Quickest Change Detection for Hidden {M}arkov Models.'' \emph{IEEE Transition of Information Theory} 65 (1): 511--529.
  
  \bibitem[Girshick and Rubin(1952)]{GR1952}
Girshick, M. A. and H. Rubin. 1952. ``A Bayes Approach to a Quality Control Model''. \emph{Annals of Mathematical Statistics} 23 (1): 114--125.

\bibitem[Kiefer and Weiss(1957)]{Kiefer&Weiss-AMS1957}
Kiefer, J., and L. Weiss. 1957. ``Properties of Generalized Sequential
  Probability Ratio Tests.'' \emph{Annals of Mathematical Statistics} 28 (1):
  57--74.
  
  \bibitem[Lai(1973)]{Lai-AS73}
Lai, T.~L. 1973.
``Optimal Stopping and Sequential Tests Which Minimize the Maximum Expected Sample Size.''
{\em Annals of Statistics} 1: 659--673.

\bibitem[Lai(1981)]{Lai-as81-SPRT}
Lai, T.~L. 1981. ``Asymptotic Optimality of Invariant Sequential
  Probability Ratio Tests.'' \emph{Annals of Statistics} 9 (2): 318--333.
 
\bibitem[Lai(1988)]{Lai-AS1988}
Lai, T.~L. 1988.
``Nearly Optimal Sequential Tests of Composite Hypotheses.''
\emph{Annals of Statistics} 16:  856--886.

\bibitem[Lai and Zhang(1994)]{LaiZhang-SQA1994}
Lai, T.~L., and L. Zhang. 1994.
``A Modification of Schwarz's Sequential Likelihood Ratio Tests in Mulitvariate Sequential Analysis.''
\emph{Sequential Analysis} 13:  79--96.

\bibitem[Lai(1998)]{LaiIEEE98}
Lai, T.~L. 1998. ``Information Bounds and Quick Detection of Parameter
  Changes in Stochastic Systems.'' \emph{IEEE Transactions on Information
  Theory} 44 (7): 2917--2929.
  
  \bibitem[Lai(2000)]{LaiIEEE00}
Lai, T.~L. 2000. ``Sequential Multiple Hypothesis Testing and Efficient Fault
  Detection-Isolation in Stochastic Systems. {\em IEEE Transactions on Information Theory}, 46(2):595--608.
  
\bibitem[Lai and Shih(2004)]{2004power}  
Lai, Tze Leung, and Mei-Chiung Shih. 2004. ``Power, Sample Size and Adaptation Considerations in the Design of Group Sequential Clinical Trials.'' 
\emph{Biometrika} 91(3): 507-528.
 
\bibitem[Lai, Liu, and Xing(2009)Lai, Liu, and Xing]{LaiLiuXing2009}
Lai, T.~L., T.~Liu, and H.~Xing. 2009.
\newblock ``A Bayesian Approach to Sequential Surveillance in Exponential
  Families."
\newblock \emph{Communications in Statistics, Theory and Methods} 38 (16):  2958--2968.

\bibitem[Lai and Siegmund(1977)]{LaiSiegmund1977} 
Lai, T. L. and D. Siegmund. 1977. ``A Nonlinear Renewal Theory With Applications to Sequential Analysis {I}.'' \emph{The Annals of Statistics} 5 (5): 946--954.

\bibitem[Lai and Siegmund(1979)]{LaiSiegmund1979} 
Lai, T. and D. Siegmund. 1979. ``A Nonlinear Renewal Theory With Applications to Sequential Analysis {II}.'' \emph{The Annals of Statistics} 5 (1): 60--76.
  
\bibitem[Lai and Xing(2008{\natexlab{a}})]{LaiXing2008}
Lai, T.~L. and H.~Xing. 2008a.
\newblock \emph{Statistical Models and Methods for Financial Markets}.
\newblock New York, USA: Springer.

\bibitem[Lai and Xing(2008{\natexlab{b}})]{LaiXing2008b}
Lai, T.~L. and H.~Xing. 2008b.
\newblock ``A Hidden Markov Filtering Approach to Multiple Change-Point Models."
\newblock In \emph{2008 47th IEEE Conference on Decision and Control}, pages
  1914--1919, Cancun, Mexico.
\newblock \doi{10.1109/CDC.2008.4739184}.

 \bibitem[Lai and Xing(2010)]{LaiXing2010}
 Lai, T.~L. and H.~Xing. 2010.
\newblock ``Sequential Change-Point Detection when the Pre- and Post-change
  Parameters are Unknown."
\newblock \emph{Sequential Analysis} 29 (2): 162--175.

\bibitem[Lai and Xing(2011)]{LaiXing2011}
 Lai, T.~L. and H.~Xing. 2011.
\newblock ``A Simple Bayesian Approach to Multiple Change-Points."
\newblock \emph{Statistica Sinica}, 21 (2): 539--569.

\bibitem[Lai and Xing(2013)]{LaiXing2013}
 Lai, T.~L. and H.~Xing. 2013.
\newblock ``Stochastic Change-Point ARX-GARCH Models and Their Applications to
  Econometric Time Series."
\newblock \emph{Statistica Sinica}, 23 (4): 1573--1594.

\bibitem[Lai et~al.(2008)Lai, Xing, and Zhang]{LaiXingZhang2008}
 Lai, T.~L., H.~Xing, and N.~Zhang. 2008.
\newblock ``Stochastic Segmentation Models for Array-Based Comparative Genomic Hybridization Data Analysis.''
\newblock \emph{Biostatistics}, 9 (2): 290--307.

\bibitem[Lai et~al., 2012]{Lai12}
Lai, T.~L., Lavori, P.~W., and Shih, M.-C. (2012).
\newblock Adaptive trial designs.
\newblock {\em Annual Review of Pharmacology and Toxicology}, 52(1):101--110.

\bibitem[Lai et~al., 2013]{Lai13}
Lai, T.~L., Liao, O. Y.-W., and Kim, D.~W. (2013).
\newblock Group sequential designs for developing and testing biomarker-guided
  personalized therapies in comparative effectiveness research.
\newblock {\em Contemporary Clinical Trials}, 36(2):651--663.

\bibitem[Lorden(1971)]{lorden-ams71}
Lorden, G. 1971. ``Procedures for Reacting to a Change in Distribution.''
  \emph{Annals of Mathematical Statistics} 42 (6): 1897--1908.
 

\bibitem[Lorden(1976)]{lorden-as76}
Lorden, G. 1976. ``{2-SPRT's} and the Modified {Kiefer-Weiss} Problem of
  Minimizing an Expected Sample Size.'' \emph{Annals of Statistics} 4 (2):
  281--291.
  
\bibitem[Lorden(1977{\natexlab{b}})]{Lorden-unpublished-1977}
Lorden, G. 1977{\natexlab{b}}. ``Nearly Optimal Sequential Tests for
  Exponential Families.'' \emph{Unpublished Manuscript}. Available from \url{http://jaybartroff.com/research/gary.pdf}

\bibitem[Lorden(1980)]{lorden-ptrf80}
Lorden, G. 1980. ``Structure of Sequential Tests Minimizing an Expected
  Sample Size.'' \emph{Probability Theory and Related Fields} 51 (2): 291--302.

\bibitem[Lorden(1983)]{Lorden83}
Lorden, Gary. 1983. ``Asymptotic efficiency of three-stage hypothesis tests.''
  \emph{The Annals of Statistics} 11: 129--140.


\bibitem[Meyn and Tweedie(2009)]{MeynTweedie2009}
Meyn, S. P. and R. L. Tweedie. 2009. \emph{Markov Chains Stochastic Stability.} 2nd
ed. Springer-Verlag, New York, USA.

\bibitem[Moustakides(1986)]{MoustakidesAS86}
Moustakides, G.~V. 1986. ``Optimal Stopping Times for Detecting Changes in
  Distributions.'' \emph{Annals of Statistics} 14 (4): 1379--1387.

\bibitem[Moustakides, Polunchenko, and Tartakovsky(2009)]{MoustPolTarCS09}
Moustakides, G.~V., A.~S. Polunchenko, and A.~G. Tartakovsky.
  2009. ``Numerical Comparison of {CUSUM and Shiryaev--Roberts} Procedures for
  Detecting Changes in Distributions.'' \emph{Communications in Statistics -
  Theory and Methods} 38 (16--17): 3225--3239.

\bibitem[Narasimhan et~al., 2022]{Narasimhan22}
Narasimhan, B., Shih, M.-C., and He, P. (2022).
\newblock {\em sp23design: Design and Simulation of Seamless Phase II-III
  Clinical Trials}.
\newblock R package version 0.9-1.

\bibitem[Nikoforov(1995)]{NikiforovIEEEIT95}
Nikiforov, I. V. 1995. ``A Generalized Change Detection Problem.''
{\em IEEE Transactions on Information Theory} 41(1):171--187.

\bibitem[Nikoforov(2000)]{NikiforovIEEEIT00}
Nikiforov, I. V. 2000. ``A Simple Recursive Algorithm for Diagnosis of Abrupt Changes in
  Random Signals.'' {\em IEEE Transactions on Information Theory} 46(7):2740--2746.
  
  \bibitem[Nikoforov(2003)]{NikiforovIEEEIT03}
Nikiforov, I. V. 2003. ``A Lower Bound for the Detection/Isolation Delay in a Class of
Sequential Tests.'' {\em IEEE Transactions on Information Theory} 49(11):3037--3046.

\bibitem[Novak(2011)]{Novak11}
Novak, S.~Y. 2011. \emph{Extreme Value Methods with Applications to
  Finance}. CRC Press.

\bibitem[Page(1954)]{page-bka54}
Page, E.~S. 1954. ``Continuous Inspection Schemes.'' \emph{Biometrika} 41
  (1--2): 100--114.
  
\bibitem[Pollak(1985)]{PollakAS85}
Pollak, M. 1985. ``Optimal Detection of a Change in Distribution.''
  \emph{Annals of Statistics} 13 (1): 206--227.

\bibitem[Schwarz(1962)]{Schwarz-AMS1962}
Schwarz, G. 1962. ``{Asymptotic Shapes of Bayes Sequential Testing Regions}.''
  \emph{Annals of Mathematical Statistics} 33 (1): 224--236.

\bibitem[Shewhart(1931)]{shewhart-book31}
Shewhart, Walter~Andrew. 1931. \emph{Economic Control of Quality of
  Manufactured Products}. New York, USA: D. Van Nostrand Co.

  \bibitem[Shih et~al., 2015]{Shih15}
Shih, M.-C., Turakhia, M., and Lai, T.~L. (2015).
\newblock Innovative designs of point-of-care comparative effectiveness trials.
\newblock {\em Contemporary Clinical Trials}, 45:61--68.

  \bibitem[Shiryaev(1963)]{shiryaev-tpa1963}
Shiryaev, A.~N. 1963. ``On Optimum Methods in Quickest Detection Problems.'' \emph{Theory of Probability and its Applications} 8 (1): 22--46.

\bibitem[Shiryaev(1969)]{shiryaev-book1969}
Shiryaev, A.~N. 1969. \emph{Statistical Sequential Analysis: Optimal Stopping Rules}. Moscow, USSR: Nauka.

\bibitem[Shiryaev(1978)]{Shiryaev1978}
Shiryaev, A.~N. 1978. \emph{Optimal Stopping Rules}.
\newblock New York, USA: Springer-Verlag.  

\bibitem[Siegmund(1985)]{siegmund-book85}
Siegmund, David. 1985. \emph{Sequential Analysis: Tests and Confidence
  Intervals}. Series in Statistics. New York, USA: Springer-Verlag.


\bibitem[Tartakovsky(1998)]{TartakovskySISP98}
Tartakovsky, A.~G. 1998. ``Asymptotic Optimality of Certain Multihypothesis
  Sequential Tests: Non-i.i.d. Case.'' \emph{Statistical Inference for
  Stochastic Processes} 1(3): 265--295.
  
  \bibitem[Tartakovsky(2008)]{Tartakovsky-SQA08b}
Tartakovsky, A. G. 2008. ``Multidecision Quickest Change-Point Detection: Previous Achievements
  and Open Problems.'' {\em Sequential Analysis} 27(2):201--231.
  
  \bibitem[Tartakovsky(2017)]{TarIEEE2017}
  Tartakovsky, A. G. 2017. ``On Asymptotic Optimality in Sequential Changepoint Detection: Non-iid Case.''
  \emph{IEEE Transactions on Information Theory} 63(6):3433--3450.
 
  
  \bibitem[Tartakovsky(2020)]{Tartakovsky_book2020}
Tartakovsky, A.~G. 2020. \emph{Sequential Change Detection and Hypothesis
  Testing: General Non-i.i.d. Stochastic Models and Asymptotically Optimal
  Rules. Monographs on Statistics and Applied Probability 165}. Boca Raton,
  London, New York: Chapman \& Hall/CRC Press, Taylor \& Francis Group.
  
  \bibitem[Tartakovsky(2025)]{TartakovskyAMSA2025}
  Tartakovsky, A.G. 2025. ``Nearly Optimum Properties of Certain Multi-Decision Sequential Rules for
  General Non-i.i.d. Stochastic Models.'' \emph{Annals of Mathematical Sciences and Applications} 10(2): 307--360. 
 
\bibitem[Tartakovsky, Nikiforov, and Basseville(2015)]{TNB_book2014}
Tartakovsky, A.~G., I.~V. Nikiforov, and M.~Basseville. 2015. \emph{Sequential
  Analysis: Hypothesis Testing and Changepoint Detection. Monographs on
  Statistics and Applied Probability 136.} Boca Raton, London, New York: Chapman
  \& Hall/CRC Press, Taylor \& Francis Group.

\bibitem[Tartakovsky, Pollak, and
  Polunchenko(2012)]{tartakovskypolpolunch-tpa11}
Tartakovsky, A. G., M. Pollak, and A.~S. Polunchenko. 2012.
  ``Third-Order Asymptotic Optimality of the Generalized {Shiryaev--Roberts}
  Changepoint Detection Procedures.'' \emph{Theory of Probability and its
  Applications} 56 (3): 457--484.
  
  \bibitem[Tartakovsky and Veeravalli(2005)]{TarVeerIEEE2005}
  Tartakovsky, A. G. and V. V. Veeravally. 2005. ``General Asymptotic Bayesian Theory of Change Detection.''
  \emph{Theory of Probability and its Applications} 49(3): 458--497.
  
\bibitem[Wald(1945)]{wald45}
Wald, Abraham. 1945. ``Sequential Tests of Statistical Hypotheses.''
  \emph{Annals of Mathematical Statistics} 16 (2): 117--186.

\bibitem[Wald(1947)]{wald47}
Wald, Abraham. 1947. \emph{Sequential Analysis}. New York, USA: John Wiley \&
  Sons, Inc.

\bibitem[Wald and Wolfowitz(1948)]{wald48}
Wald, Abraham, and J.~Wolfowitz. 1948. ``Optimum Character of the Sequential
  Probability Ratio Test.'' \emph{Annals of Mathematical Statistics} 19 (3):
  326--339.

\bibitem[Woodroofe(1976)]{Woodroofe1976}
Woodroofe, M. 1976. ``A Renewal Theorem for Curved Boundaries and Moments of First Passage Tmes.'' \emph{The Annals of Probability} 4 (1): 67--80.

\bibitem[Woodroofe(1977)]{Woodroofe1977}
Woodroofe, M. 1977. ``Second Order Approximations for Sequential Point and Interval Estimation.'' \emph{The Annals of Statistics} 5 (5): 984--995.

\bibitem[Woodroofe(1982)]{Woodroofe1982} 
Woodroofe, M. 1982. \emph{Nonlinear Renewal Theory in Sequential Analysis.}  Philadelphia, USA: SIAM.
 
\bibitem[Xing et~al.(2012{\natexlab{a}})Xing, Mo, Liao, and
  Zhang]{XingEtAl2012b}
Xing, H., Y.~Mo, W.~Liao, and M.~Zhang. 2012a.
\newblock ``Genomewide Localization of Protein-dna Binding and Histone
  Modification by BCP with Chip-seq Data."
\newblock \emph{PLoS Computational Biology}, 8 (7): e1002613.
\newblock \doi{10.1371/journal.pcbi.1002613}.

\bibitem[Xing et~al.(2012{\natexlab{b}})Xing, Sun, and Chen]{XingEtAl2012}
Xing, H., N.~Sun, and Y.~Chen. 2012b.
\newblock ``Credit Rating Dynamics in the Presence of Unknown Structural Breaks."
\newblock \emph{Journal of Banking and Finance}, 36 (1): 
78--89.

\bibitem[Xing et~al.(2020)Xing, Wang, Li, and Chen]{XingEtAl2020}
Xing, H., K.~Wang, Z.~Li, and Y.~Chen. 2020.
\newblock ``Statistical Surveillance of Structural Breaks 
in Credit Rating Dynamics." 
\newblock \emph{Entropy}, 22: 1072.
\newblock \doi{10.3390/e22101072.}

\bibitem[Zacks(1991)]{Zacks1991}
Zacks, S. 1991. ``Detection and Change-Point Problems."
\newblock \emph{Handbook of Sequential Analysis}, edited 
by B.~K. Ghosh and P.~K. San.
531--562. New York, USA: Marcel Dekker.

\bibitem[Zacks and Barzily(1981)]{ZacksBarzily1981}
Zacks, S. and Z.~Barzily. 1981. 
\newblock ``Bayes Procedures for Detecting a Shift in the Pprobability of Success
  in a Series of Bernoulli Trials."
\newblock \emph{Journal of Statistical Planning and Inference} 5: 107--119.

\bibitem[Zhang(1988)]{Zhang1988}
 Zhang, C.-H. 1988. ``A Nonlinear Renewal Theory'' \emph{The Annals of Probability} 16 (2) 793--824.


\end{thebibliography}

\end{document}